\documentclass[fleqn,usenatbib]{mnras}

\usepackage{newtxtext,newtxmath}
\usepackage[T1]{fontenc}

\DeclareRobustCommand{\VAN}[3]{#2}
\let\VANthebibliography\thebibliography
\def\thebibliography{\DeclareRobustCommand{\VAN}[3]{##3}\VANthebibliography}

\usepackage{graphicx}	% Including figure files
\usepackage{amsmath}	% Advanced maths commands
\usepackage{amssymb}	% Extra maths symbols
\usepackage{gensymb}

\usepackage{symbols}
\usepackage{enumitem}
\usepackage{footnotehyper}
\usepackage{hyperref}
\usepackage{siunitx}

\title[The WISDOM of power spectra]{The WISDOM of power spectra: how the galactic gravitational potential impacts a galaxy's central gas reservoir in simulations and observations}

\author[J. Gensior et al.]{
\parbox{\textwidth}{Jindra Gensior,$^{1,2}$\thanks{E-mail: jindra.gensior@uzh.ch}
Timothy A. Davis,$^{3}$
Martin Bureau,$^{4}$
J.~M.~Diederik Kruijssen,$^{5,6}$
Michele Cappellari,$^{4}$
Ilaria Ruffa,$^{3}$ and
Thomas G. Williams$^{4}$
}
\\
$^{1}$Institute for Computational Science, University of Zurich, Winterthurerstrasse 190, 8057 Z{\"u}rich, Switzerland\\
$^{2}$Astronomisches Rechen-Institut, Zentrum f{\"u}r Astronomie der Universit{\"a}t Heidelberg, M{\"o}nchhofstra{\ss}e 12-14, 69120 Heidelberg, Germany\\
$^{3}$Cardiff Hub for Astrophysics Research \&\ Technology, School of Physics \&\ Astronomy, Cardiff University, Queens Buildings, Cardiff, CF24 3AA, UK\\
$^{4}$Sub-department of Astrophysics, Department of Physics, University of Oxford, Keble Road, Oxford OX1 3RH, UK\\
$^{5}$Technical University of Munich, School of Engineering and Design, Department of Aerospace and Geodesy, Chair of Remote Sensing Technology, \\\hspace{2.2mm}Arcisstr. 21, 80333 Munich, Germany\\
$^{6}$Cosmic Origins Of Life (COOL) Research DAO, coolresearch.io
}

\date{Accepted 2023 October 09. Received 2023 September 13; in original form 2023 February 03}

\pubyear{2023}

\begin{document}
\label{firstpage}
\pagerange{\pageref{firstpage}--\pageref{lastpage}}
\maketitle

\begin{abstract}
{Observations indicate that the central gas discs are smoother in early-type galaxies than their late-type counterparts, while recent simulations predict that the dynamical suppression of star formation in spheroid-dominated galaxies is preceded by the suppression of fragmentation of their interstellar media. The mass surface density power spectrum is a powerful tool to constrain the degree of structure within a gas reservoir. Specifically here, we focus on the power spectrum slope and aim to constrain whether the shear induced by a dominant spheroidal potential can induce sufficient turbulence to suppress fragmentation, resulting in the smooth central gas discs observed. We compute surface density power spectra for the nuclear gas reservoirs of fourteen simulated isolated galaxies and twelve galaxies observed as part of the mm-Wave Interferometric Survey of Dark Object Masses (WISDOM) project. Both simulated and observed galaxies range from disc-dominated galaxies to spheroids, with central stellar mass surface densities, a measure of bulge dominance, varying by more than an order of magnitude. For the simulations, the power spectra steepen with increasing central stellar mass surface density, thereby clearly linking the suppression of fragmentation to the shear-driven turbulence induced by the spheroid. The WISDOM observations show a different (but potentially consistent) picture: while there is no correlation between the power spectrum slopes and the central stellar mass surface densities, the slopes scatter around a value of 2.6. This is similar to the behaviour of the slopes of the simulated galaxies with high central stellar mass surface densities, and could indicate that high shear eventually drives incompressible turbulence.}
\end{abstract}

\begin{keywords}
galaxies: elliptical and lenticular, cD -- galaxies: ISM -- galaxies: spiral -- galaxies: structure -- ISM: structure
\end{keywords}

%%%%%%%%%%%%%%%%%%%%%%%%%%%%%%%%%%%%%%%%%%%%%%%%%%

%%%%%%%%%%%%%%%%% BODY OF PAPER %%%%%%%%%%%%%%%%%%

\section{Introduction} \label{s:Intro}
How galaxies become quiescent and cease their star formation activity despite the presence of molecular gas reservoirs, remains an open question. Empirically, the total star formation rates (SFRs) of late-type, star-forming galaxies are proportional to the mass surface densities of their gas reservoirs \citep[e.g.][]{Kennicutt1998,Bigiel2008,delosReyes2019}. By contrast, the $\approx25$~per~cent of early-type galaxies (ETGs) which host molecular gas \citep[e.g.][]{Young2011,Davis2019} exhibit lower SFRs despite hosting gas reservoirs with similar mass surface densities \citep{Davis2014}. A similar behaviour is seen in the central, bulge-dominated region of the Milky Way \citep{Longmore2013,Kruijssen2014}. Additionally, there is some evidence of intermediate-redshift ETGs that are quiescent despite hosting molecular gas reservoirs \citep{Suess2017,Williams2021}.

In recent years, the `dynamical suppression' of star formation \citep[also known as `morphological quenching'\footnote{For the remainder of this paper we will exclusively use the term \textit{dynamical suppression}, since the galactic dynamics play a crucial role in the suppression and quenching, whereas morphological quenching places all the emphasis on the presence of a bulge.};][]{Martig2009} has emerged as a viable pathway to quiescence in bulge-dominated galaxies \citep[e.g.][]{Martig2009,Martig2013,Gensior2020,Gensior2021,Kretschmer2020}. Dynamical suppression describes the process whereby shear, induced by the spheroidal component of the gravitational potential, drives turbulence in the gas, thereby stabilising it against collapse and suppressing star formation. Although dynamical suppression requires low gas fractions to be effective \citep{Martig2013,Gensior2021}, it offers a compelling explanation for the suppressed SFRs observed in ETGs with gas reservoirs. Furthermore, dynamical suppression is predicted to maintain quiescence in a spheroid-dominated galaxy when its cold interstellar medium (ISM) is replenished through gas-rich minor mergers and stellar mass loss \citep[e.g.][]{Davis2019}.

A key step for the dynamical suppression of star formation is the (dynamical) suppression of fragmentation. \citet{Gensior2020} simulated ten isolated galaxies with an identical stellar mass, but morphologically ranging from pure disc to spheroid. The galaxies with dominant bulges formed smooth, dense circumnuclear gas discs that increased in size with increasing central stellar mass surface density ($\mus$, a proxy for the dominance of the spheroidal component). By comparison, the low-$\mus$ galaxies in the \citet{Gensior2020} sample host ISM that are porous, sub-structured and clumpy throughout. \citet{Davis2022} used non-parametric morphological indicators to quantify the cold ISM structure of the galaxies from \citet{Gensior2020}, the Physics at High Angular resolution in Nearby GalaxieS (PHANGS) survey \citep[e.g][]{Leroy2021b,Leroy2021a} and the mm-Wave Interferometric Survey of Dark Object Masses (WISDOM)\footnote{\url{https://wisdom-project.org/}} sample. They found strong anti-correlations between the central stellar surface density and the Gini, Smoothness and Asymmetry coefficients in both simulations and observations, demonstrating that galaxies with a deep central gravitational potential (high $\mus$) have a more smooth, symmetric and regular cold ISM (low Gini, Asymmetry, Smoothness).
In this paper we wish to further quantify the effects the gravitational potential and galactic dynamics have on the gas reservoir of a galaxy and its turbulent state. To do so, we turn to an ISM statistic that has been used numerous times to analyse turbulence in galaxies: the spatial power spectrum of the ISM.

The power spectrum encodes information on the nature of the turbulence \citep[e.g.][]{Elmegreen2004}. For example, the slopes of the density and velocity power spectra can be used to distinguish whether the turbulence is incompressible \citep{Kolmogorov1941}, or compressible \citep[e.g.][]{Burgers1948,Fleck1996}, and whether the turbulent forcing is solenoidal or compressive \citep[e.g.][for an application to a spiral galaxy]{Federrath2013,Nandakumar2020}. Power spectra have been extensively studied within regions of the Milky Way \citep[e.g.][]{Crovisier1983,Green1993,Stutzki1998,Lazarian2000,Miville-Deschenes2003,Swift2008,Miville-Deschenes2010,Martin2015,Pingel2018}. Over the last two decades this has been extended to external galaxies such as the Local Group (e.g. \citealt{Stanimirovic1999} for the Small Magellanic Cloud and \citealt{Elmegreen2001},\citealt{Block2010} and \citealt{Colman2022} for the Large Magellanic Cloud, see also \citealt{Koch2020} for both Magellanic Clouds, M31 and M33) and nearby spirals \citep[e.g.][]{Dutta2013} and dwarfs \citep[e.g.][]{Dutta2009b,Zhang2012}. 

Crucially for our purpose, the density power spectrum is very sensitive to the morphology of the tracer \citep{Koch2020}. Specifically, the absence of fragmentation shifts power from the smaller to the larger spatial scales \citep{Renaud2013, Grisdale2017}. Therefore, a suppression of fragmentation in the data will manifest itself as a steep power spectrum. However, there is some debate surrounding the shape (and slope) of extragalactic power spectra. Some studies find that their data are better represented by a broken power-law \citep[e.g.][]{Dutta2008,Dutta2009b,Block2010,Combes2012}, where the small-scale power-law has a steeper slope than the large-scale one. This break in the power spectrum is hypothesised to indicate the transition from three-dimensional turbulence on small scales to two-dimensional turbulence on large scales, the location of the break coinciding with the scale height of the galaxy's gas disc \citep[e.g.][]{Dutta2008}. Alternatively, based on numerical simulations, \citet{Koertgen2021} proposed that the break scale in cold gas traces the average size of molecular clouds, which is not necessarily equal to the disc scale height. Yet, a break in the power spectrum is far from ubiquitous. \citep[e.g.][]{Stanimirovic1999,Dutta2009a,Zhang2012,Dutta2013,Koch2020}. Recently, \citet{Koch2020} found evidence that the break could be an artefact from not accounting for the shape of the point spread function of the data and can otherwise result from a bright, dominant source \citep[see also][]{Willet2005}. Similarly, different methods of calculating the power spectrum (in particular, whether the \emph{uv}-plane visibility or the zeroth-moment, mass surface density map is used in the computation\footnote{For example, the measured slope for NGC~628 ranges from 1.6 \citep[][visibility method]{Dutta2008}, to 2.2 \citep[][surface density map]{Walker2014}, to 2.6 \citep[][surface density map]{Grisdale2017}. See also the discussion concerning DDO 210 in \citet{Zhang2012}.}) and different tracers (e.g. \citealt{Combes2012, Koch2020} and references therein) can yield very different power spectrum slopes. The above could imply that the density power spectrum is a poorly constrained probe of ISM turbulence, at least when considering whole galaxies. However, given its sensitivity to the morphology of the gas, and given the ability to systematically apply the same procedure to the data, the density power spectrum should be a good tool to quantify the impact of the gravitational potential on the turbulence within a galaxy. If the smooth, circumnuclear gas reservoirs of bulge-dominated galaxies are the result of shear-driven turbulence, we expect a continuous steepening of the power spectrum slope going from disc-dominated galaxies with a porous and sub-structured ISM to bulge-dominated ones. 

Given the good agreement between non-parametric morphological indicators (and their trends with galaxy type) between the \citet{Gensior2020} simulated galaxies and observed galaxies in the local universe \citep[see][]{Davis2022}, we wish to compare the power spectra findings of the simulations to observations. To do so, we turn to the WISDOM project \citep[e.g.][]{Onishi2017_WISDOMI_NGC3665,North2019_NGC0383,Smith2019_NGC0524,Liu2021,Davis2022}, which akin to the simulations includes galaxies with diverse stellar morphologies. The primary aim of WISDOM is to accurately measure the mass of each galaxy's central super-massive black hole (SMBH) using the molecular gas dynamics. As a result, exquisite, high-resolution CO maps of the circumnuclear gas reservoirs are available for a variety of galaxies, ranging from late to early types. This makes the WISDOM sample the ideal comparison sample for our purposes. 

This paper is structured as follows. In Section~\ref{s:Methods} we give an overview of the \citet{Gensior2020} simulations and the WISDOM observations used, as well as of our method to measure the density power spectrum. The power spectra of simulations and observations are presented in Section~\ref{s:PS} and their trends are discussed in Section~\ref{s:Disc}. We summarise our findings and conclude in Section~\ref{s:Conclusion}.

\section{Methods}\label{s:Methods}
\subsection{Simulations}\label{ss:Sims}
\subsubsection{Simulations set-up}\label{sss:simICs}
We analyse the ISM morphologies of the isolated galaxies presented in \citet{Gensior2020}, that were simulated with a dynamics-dependent star formation efficiency (SFE), as well as the ISM morphology of four additional simulations. We briefly describe the initial conditions and simulation physics below, but refer the reader to \citet{Gensior2020} for a more complete description.

The \citet{Gensior2020} suite comprises ten simulations, where each isolated galaxy consists of a gaseous and a stellar component within a \citet{Hernquist1990} dark matter halo. In all cases excepting one (run noB), the stellar contribution has been split into an exponential disc and a \citet{Hernquist1990} bulge component. The total initial stellar mass is fixed at $\Mstar \approx 4.71 \times 10^{10}~\Msun$, but the initial mass fraction of the bulge component, $\Mb$, is varied from 30 to 90~per~cent, thereby varying the bulge-to-disc and bulge-to-total mass ratio between the simulations. Simultaneously the initial bulge scale radius, $\Rb$, is varied from 1 to 3 kpc across the simulations to create a range of different gravitational potentials. All galaxies are initialised with a gas-to-stellar mass ratio $M_{\rm gas}/\Mstar = 0.05$, in good agreement with the cold gas fraction of massive galaxies in the local Universe \citep[e.g.][]{Saintonge2017}, and within the upper range of gas fractions of early-type galaxies \citep[][]{Young2011}. 

The central stellar mass surface density is defined as
\begin{equation}\label{eq:III_mus}
    \mus \equiv \frac{\Mstar}{2\pi \Reff^2},
\end{equation}
where $\Reff$ is the effective radius. For the simulations, we use the stellar half-mass radius, i.e. the radius in the $x$-$y$ plane of the galaxy which contains 50~per~cent of the total stellar mass, to compute $\mus$. Similar to the gas fraction and bulge-to-disc ratio, the central stellar mass surface density changes negligibly during runtime. Hence, we refer to all simulated galaxies with their initial conditions.
Due to the sparsity of simulated galaxies with $\log (\mus/\Msun\kpc^{-2}) \gtrsim 8.75$ in \citet{Gensior2020} compared to the WISDOM galaxies in our sample, we extend the \citet{Gensior2020} suite by performing four additional simulations. Initial conditions for the additional simulations were created following \citet{Springel2005} as outlined in the previous paragraph, again only changing the distribution of the stellar contribution to the potential. To add more high-$\mus$ galaxies, we include one intermediate mass, compact galaxy ($\Mb=0.75\Mstar$, $\Rb=1$~kpc), an additional $\Mb=0.9\Mstar$ galaxy with $\Rb=1.5$~kpc, and two pure spheroids ($\Mb=\Mstar$) with scale radii of 1 and 2 kpc respectively. We summarise the properties of our initial conditions in Table ~\ref{tab:simIC_properties}.

\begin{table}
 \caption{Initial conditions of the simulations.}
 \begin{tabular}{lcccc}
  \hline
  Name & M$_{\rm b}$  & B/D & R$_{\rm b}$ & $\log (\mus)$ \\
  & (10$^{10}\Msun$) & & (kpc) & $(\Msun \kpc^{-2})$\\
  \hline
  noB    &  0.00 &  0.00 & 0 & 8.09 \\
  B\_M30\_R1 & 1.41 & 0.43 & 1 & 8.33 \\
  B\_M30\_R2 & 1.41 & 0.43 & 2 &  8.24\\
  B\_M30\_R3 & 1.41 & 0.43 & 3 & 8.17\\  
  B\_M60\_R1 & 2.83 & 1.50 & 1 & 8.74 \\
  B\_M60\_R2 & 2.83 & 1.50 & 2 & 8.46 \\
  B\_M60\_R3 & 2.83 & 1.50 & 3 & 8.30 \\
  B\_M90\_R1 & 4.24 & 9.00 & 1 & 9.28 \\
  B\_M90\_R2 & 4.24 & 9.00 & 2 & 8.76 \\
  B\_M90\_R3 & 4.24 & 9.00 & 3 & 8.49 \\  
  \hline
  B\_M75\_R1 & 3.50 & 3.00 & 1 & 9.01 \\
  B\_M90\_R1.5 & 4.24 & 9.00  & 1.5 & 8.96 \\
  B\_M100\_R1 & 4.67 & $\infty$ & 1 & 9.43  \\  
  B\_M100\_R2 & 4.67 & $\infty$ & 2 & 8.86 \\
  \hline
  \multicolumn{5}{p{0.95\linewidth}}{\footnotesize{\textit{Notes:} The top ten simulations were presented in \citet{Gensior2020}, while the latter four have been added here. Column 1 lists the name of each simulation, following the naming convention of \citet{Gensior2020}: the prefix refers to the presence of a bulge, the subsequent `MX' and `RY' to the bulge mass fraction in percent and the bulge scale radius in kpc, respectively. Column 2 lists the initial bulge mass, column 3 lists the bulge-to-disc stellar mass ratio, column 4 lists the bulge scale radius and column 5 lists the central stellar mass surface density $\mus$.}}\\
 \end{tabular}
 \label{tab:simIC_properties}
\end{table}

The simulations are performed with the moving-mesh code {\sc{Arepo}} \citep{Springel2010,Weinberger2020}, which treats stars and dark matter as Lagrangian particles, while the hydrodynamics are solved on an unstructured mesh in form of a Voronoi tessellation. Gravity is solved with a tree-based scheme and we use an adaptive gravitational softening approach \citep{Price2007} for optimal gravitational resolution. With a mass resolution of $\sim 10^4~\Msun$ for gas and stars, and a mass resolution of $\sim 10^5~\Msun$ for the dark matter, we fix the minimum softening length at 12 pc for gas and stars and at 26 pc for dark matter, respectively. This ensures that the gas is well resolved spatially at our star formation density threshold $\rm n_{\rm thresh} = 1 ~\ccm$ and higher. 

Star formation is modelled using the \citet{Katz1992} parametrisation for the volumetric SFR density $\dot{\rho}_{\rm SFR} = \eff\rho\tff^{-1}$, where $\eff$ is the SFE per free-fall time, $\rho$ the mass volume density and $\tff = \sqrt{3\pi/32G\rho}$ the free-fall time of the gas. We introduce a dependence on the gas dynamics, using the virial parameter, $\avir$, dependent SFE of \citet{Padoan2012,Padoan2017}:
\begin{equation}\label{eq:III_sfe}
    \eff = 0.4\exp(-1.6\avir^{0.5}).
\end{equation}
This dynamics-dependent SFE reproduces the observed suppressed SFRs of ETGs \citep{Gensior2020,Gensior2021,Kretschmer2020}, but converges to the \citet{Kennicutt1998} relation for late-type or gas-rich galaxies \citep[e.g.][]{Semenov2016,Gensior2021}. 

The virial parameter is calculated from a cloud-like overdensity around each star-forming gas cell using the algorithm presented in \citet{Gensior2020}. Apart from a minimum density threshold of $\rm n_{\rm thresh} = 1 ~\ccm$, the gas must also be cooler than $10^3$ K to be eligible for star formation. We model the thermal state of the gas using the Grackle chemistry and cooling library\footnote{\url{https://grackle.readthedocs.io/}} \citep{Smith2017} with the six-species chemistry network enabled. Tabulated atomic metal line cooling at solar metallicity and heating from the interstellar radiation field using the \citet{Haardt2012} UV-background are included as well. 

We consider feedback from massive stars in the form of type II supernovae (SNe). These are modelled using the mechanical feedback formulation \citep{Kimm2014,Hopkins2014}: numerically the energy, mass and metals injected by each SN are distributed to the 32 nearest neighbours using a smooth particle hydrodynamics kernel. We assume one SN per 100~$\Msun$ of stellar mass formed \citep{Chabrier2003,Leitherer2014}, where each SN injects $10^{51}\ergs$ of energy and ejects 10.5~$\Msun$ of mass and 2~$\Msun$ of metals into the surrounding medium. The SNe are detonated with a delay time of 4 Myr, towards the upper end of the range derived from recent observations of feedback disruption \citep{Kruijssen2019,Chevance2020c,Chevance2020b}.

\subsubsection{Obtaining gas surface density maps} \label{sss:SDmaps}

A total gas mass surface density ($\Sigma_{\rm gas}$) map of each simulation snapshot is created using the ray-tracing functionality within {\sc{AREPO}}. The primary focus of this work is to analyse the circumnuclear gas discs of galaxies, therefore we create face-on maps that focus on the central 2.25 kpc of each simulated galaxy. This is comparable in extent to the larger WISDOM maps, e.g. that of NGC 0383 (\citealt{North2019_NGC0383}, see also \citealt{Davis2022}). Each gas surface density map is computed by integrating the densities of all the gas cells that fall within the line of sight of each pixel, from 1 kpc below to 1 kpc above the galactic mid-plane. The pixel size of our original maps is 1.2 pc per pixel, comparable to the size of the densest Voronoi cells. The size is chosen to ensure that all structures are sufficiently spatially resolved. To make this more comparable to the observations, we then smooth each map with a Gaussian kernel of full width at half maximum (FWHM) of 43 pc, equal to the geometric average of the NGC 0383 synthesised beam. Following this smoothing, each map is regridded to a coarser resolution of 10.8 pc per pixel so that the pixel grid nyquist samples the convolution kernel, while conserving surface density. 

Although we use NGC 0383 as a point of reference, we tested the spatial resolution dependence of our results using different kernel sizes. We found that this size does not qualitatively affect our conclusions, because fits to the power spectra are restricted to spatial scales larger than the kernel size and we focus on the best-fitting power-law slopes during our main analysis (see Section~\ref{ss:PS} for more details on how we compute the power spectra, and Appendix~\ref{As:BS} for details of the resolution test). While the surface density maps of the simulations include contributions from all gas cells within the field of view, which will predominantly be cold and dense, the WISDOM observations only map molecular gas. Hence, we tested how our results are affected by only including gas above different mass surface density thresholds. Using a threshold of 10\,$\Msun~\pc^{-2}$, above which the gas is expected to be predominantly molecular \citep[e.g.][]{Krumholz2009}, does not alter our conclusions. Additionally, we created mock radio interferometric observations of the simulations using the KINematic Molecular Simulation (\texttt{KinMS})\footnote{\url{https://kinms.space/}} tool \citep{Davis2013b,Davis2020_KinMS}. Similarly to the resolution test, we did not find any qualitative difference between the results of the {\sc{Arepo}} projections and the KinMS maps. Therefore, we use the gas mass surface density projections as described above and refer the interested reader to Appendices~\ref{As:KinMS} and \ref{As:SD_thresh} for a more quantitative discussion of these tests.

\begin{figure*}
    \centering
    \includegraphics[width=0.95\linewidth]{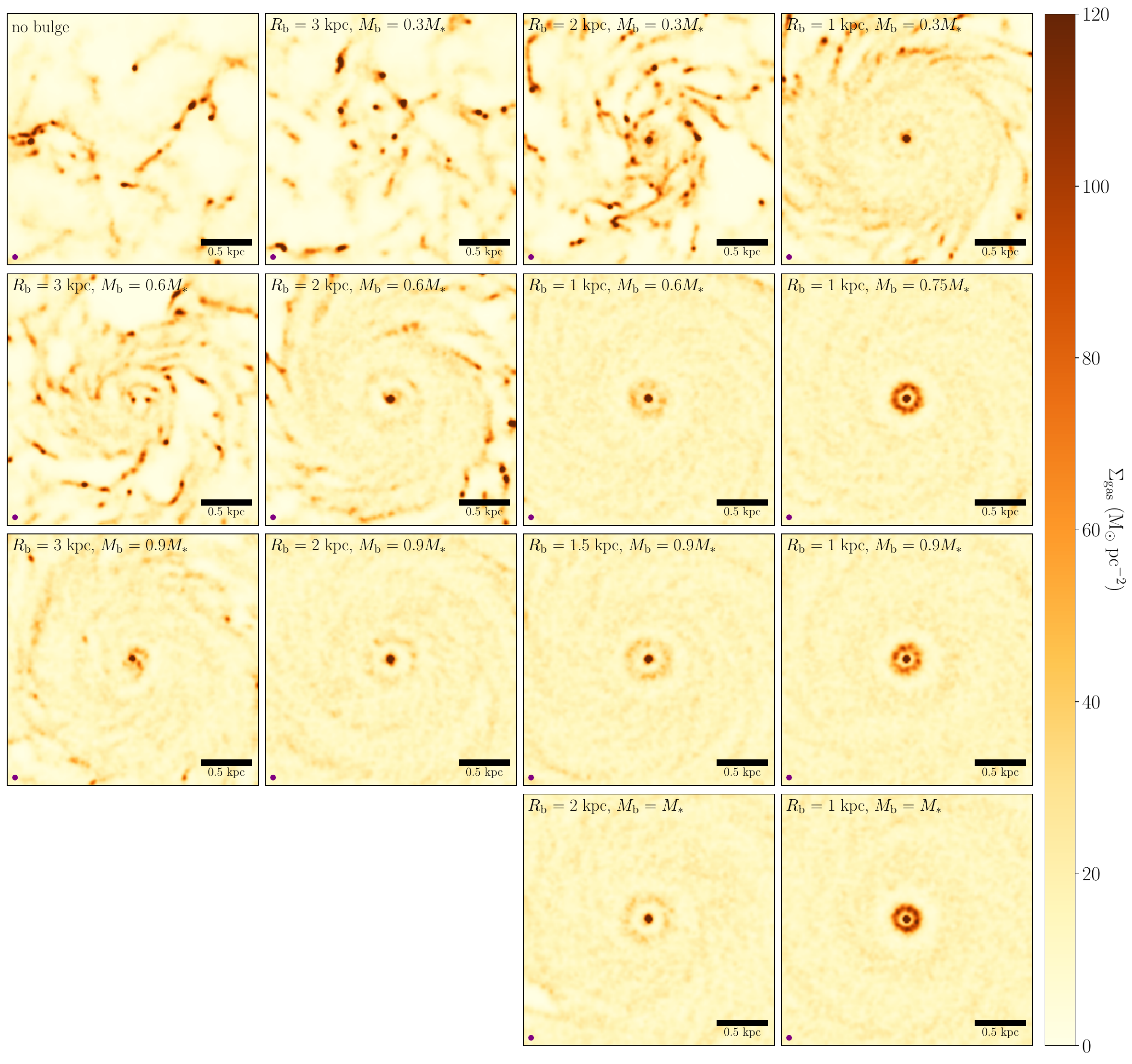}
    \caption{Gaussian-smoothed and regridded mass surface density projection of the central gas reservoir of each simulation considered in this paper, 600 Myr after the start of the simulation. The simulations are ordered by increasing central stellar mass surface density, from the lowest (noB, top-left) to highest (B\_M100\_R1, bottom-right). Each map has a linear extent of 2.25 kpc and the purple solid circle in the bottom-left corner indicates the size of the Gaussian beam. The central regions of the bulge-dominated galaxies (bottom right) are much smoother and largely devoid of sub-structure, compared to the clumpy ISM found in the central regions of the disc-dominated galaxies (top left).}
    \label{fig:gasSD_sims}
\end{figure*}

To ensure that we average over fluctuations due to cloud-scale rapid cycling of gas \citep[e.g.][]{Kruijssen2019,Chevance2020b}, we generate 8 surface density maps per simulated galaxy, starting from the 300 Myr snapshot, when the gas disc has settled into equilibrium. Subsequently a map is generated for each of seven additional snapshots separated by 100 Myr, approximately a galactic dynamical time, until the end of the simulation at 1 Gyr. Figure~\ref{fig:gasSD_sims} shows the gas mass surface density map of the central region of each simulated galaxy, 600 Myr after the beginning of the simulation. One can clearly see the aforementioned differences of ISM structure: the maps of the disc-dominated galaxies (top-left panels) are dominated by dense clumps and filamentary structures, whereas the surface density maps of the bulge-dominated galaxies (bottom-right panels) appear very smooth, apart from the very bright central mass concentrations.

\subsection{Observational Data}\label{ss:Obs}
The molecular gas mass surface density (obtained from the zeroth-moment) maps of twelve galaxies from the WISDOM project constitute the observational component of our comparison, making it the largest sample of CO extragalactic power spectra studied to date. This sample comprises five late-type and six early-type galaxies as well as one dwarf spheroidal galaxy, with a range of central stellar mass surface densities. We calculate $\mus$ following Equation~\ref{eq:III_mus}, using stellar masses derived from dynamical models where available, or from the $z$ = 0 Multiwavelength Galaxy Synthesis survey \citep[z0MGS;][]{Leroy2019} otherwise. The $K$-band effective radius is used as $\Reff$ in Equation~\ref{eq:III_mus} and taken from the Two Micron All Sky Survey \citep[2MASS;][]{Skrutskie2006} whenever possible. Table~\ref{tab:obs_props} lists the selected galaxies, their associated properties and observational references. We assume uncertainties of 0.1 dex for gas and stellar mass, ${1.5''}$ for $\Reff$ and 0.2 dex for the SFR. 

\begin{savenotes}
\begin{table*}
\caption{WISDOM galaxies and their properties}
 \begin{tabular}{lllllllllll}
  \hline
  Galaxy & Type & Distance & $\log (M_{\rm H2})$ & $\log (\Mstar)$ & R$_{\rm e, K}$ & $\log (\mus)$ & $\log ({\rm SFR})$& Inclination & PA & WISDOM \\
  & & (Mpc) & ($\Msun$) & ($\Msun$) & (arcsec) & ($\Msun$ $\kpc^{-2}$) & ($\Msun$ $\yr^{-1}$) & ($\degree$) & ($\degree$) &\\
  \hline
  FRL 0049 & -3.0 (ETG)
  & 86.7 [1] & 8.7 [1]& 10.3 [1] & 3 [1] & 9.31 & 0.78 [1] & 55.9 & 30 & 1, 9 \\
  NGC 0383 & -2.9 (ETG) & 66.6 [2] & 9.2 [9]& 11.8 [12] & 11 [15]  & 9.92 & 0.00 [9]& 37.6 & 142 & 9, 18, 19 \\
  NGC 0404 & -2.8 (ETG) & 3.06 [3] & 6.7 [10]& 9.1 [13] & 64 [16] & 8.34 & -2.60 [17] & 20.0 & 37 & 20, 21 \\
  NGC 0524 & -1.2 (ETG) & 23.3 [4] & 7.9 [9]& 11.4 [4] & 24 [15] & 9.75 & -0.56 [14] & 20.6 & 40 &  9, 19, 22 \\
  NGC 1387 & -2.8 (ETG) & 19.9 [5] & 8.3 [9]& 10.7 [14] & 16 [15] & 9.51 & -0.68 [14]& 35.0 & 64 & 9, 23 \\
  NGC 1574 & -2.9 (ETG) & 19.3 [6] & 7.5 [11]& 10.8 [14] & 21 [15] & 9.41 & -0.91 [14] & 25.0 & 340 & 9, 11, 19 \\
  NGC 4429 & -0.8 (ETG) & 16.5 [4] & 8.0 [9]& 11.2 [4] & 49 [15] & 9.19 & -0.84 [14]& 66.8 & 93 & 9, 19, 24, 25\\
  NGC 4501 & 3.3 (Spiral) & 15.3 [4] & 8.9 [9]& 11.0 [14] & 58 [15] & 8.94 & 0.43 [14] & 58.7 & 135 & 9\\
  NGC 4826 & 2.2 (Spiral) & 7.36 [4] & 7.9 [9]& 10.2 [14] & 69 [15] & 8.62 & -0.71 [14] & 59.5 & 100 &  9\\
  NGC 5806 & 3.2 (Spiral) & 21.4 [4] & 9.0 [9]& 10.6 [14] & 30 [15] & 8.80 & -0.03 [14] & 60.0 & 170 & 9, 26\\
  NGC 6753 & 3.0 (Spiral) & 43.7 [7] & 9.6 [9]& 10.8 [14] & 20 [15] & 8.78 & 0.32 [14]& 31.0 & 30 & 9\\
  NGC 6958 & -3.7 (ETG) & 35.4 [8] & 8.7 [9]& 10.8 [14] & 12 [15] & 9.35 & -0.58 [14] & 70.0 & 115 &  9, 27 \\  
  \hline
%   \hline
\multicolumn{11}{p{0.95\linewidth}}{\footnotesize{\textit{Notes:} For each galaxy, column 1 lists its name and column 2 its morphological type according to the HyperLEDA database \citep{Makarov2014}. For the following properties, a reference to the source is indicated in square brackets and listed at the end of this description. The galaxy distance is listed in column 3, the molecular gas mass, estimated from the moment zero map as described in the text, in column 4. The total stellar mass and the $K$-band effective radius are listed in columns 5 and 6, respectively. Column 7 lists the central stellar mass surface density, calculated from the stellar mass and effective radius as described in the text. Column 8 lists the SFR, column 9 the inclination and column 10 the position angle (PA) of the galaxy. Column 11 lists the associated WISDOM papers. \textit{References}: (1) \citet{Lelli2022} (2) \citet{Freedman2001}, (3) \citet{Karachentsev2002}, (4) \citet{Cappellari2011}, (5) \citet{Liu2002}, (6) \citet{Tonry2001}, (7) \citet{Bogdan2017}, (8) \citet{Marino2011}, (9) \citet{Davis2022}, (10) this work,  (11) \citet{Ruffa2023}, (12) \citet{Veale2017}, (13) \citet{Seth2010}, (14) \citet{Leroy2019}, (15) \citet{Skrutskie2006}, (16) \citet{Baggett1998}, (17) \citet{Thilker2010}, (18) \citet{North2019_NGC0383}, (19) \citet{Williams2023}, (20) \citet{Davis2020_NGC0404}, (21) \citet{Liu2022}, (22) \citet{Smith2019_NGC0524}, (23) Boyce et al. (in prep.),  (24) \citet{Davis2018_NGC4429}, (25) \citet{Liu2021}, (26) \citet{Choi2023}, (27) Thater et al. (in prep.).}} \\
 \end{tabular}
 \label{tab:obs_props}
\end{table*}
\end{savenotes}

All selected WISDOM galaxies were observed with the Atacama Large Millimetre/submillimetre Array (ALMA) in both a compact and an extended configuration, such that the data are sensitive to emission on spatial scales up to $\approx$11 arcsec in all cases. For ten galaxies the $J=2$$\rightarrow$1 transition of $^{12}\rm CO$ was observed. For NGC 4429 and NGC 4826 the $^{12}\rm CO$(3-2) line was observed instead. The raw ALMA data for each galaxy were calibrated using the standard ALMA pipeline, as provided by ALMA regional centre staff. The Common Astronomy Software Applications \citep[{\textsc{Casa}};][]{McMullin2007} package was then used to create the final data cube, from which the zeroth-moment map was generated using a masked moment technique \citep[e.g.][]{Dame2011}. A more detailed description of the data reduction and calibration can be found in the relevant, associated WISDOM project papers \citep[e.g.][and particularly \citealt{Davis2022}]{Davis2017_NGC4697,Davis2018_NGC4429,Davis2020_NGC0404,North2019_NGC0383,North2021_NGC708,Smith2019_NGC0524,Smith2021_NGC7052, Liu2021,Liu2022, Lu2022}. We use this zeroth-moment map to estimate the total molecular gas mass ($\MHt$) listed in Table~\ref{tab:obs_props}. 
To convert from brightness temperatures to surface densities we assume a CO-to-H$_2$ conversion factor $\alpha_{\rm CO}= 4.36~\Msun(\K~\kms)^{-1}\pc^{-2}$ (corresponding to $X_{\rm CO} = 2\times10^{20}~\cm^{-2} (\K~\kms)^{-1}$, \citealt{Dickman1986}), a CO(2-1)/CO(1-0) intensity ratio (in temperature units) of  0.7 and a CO(3-2)/CO(1-0) intensity ratio of 0.3 \citep{Leroy2021c}. This is the same procedure as used in the companion paper on ISM morphologies \citealt{Davis2022}. Table~\ref{tab:obs_stats} details the properties such as synthesised beam and pixel sizes for all galaxies in the sample.

\begin{table}
\caption{Zeroth-moment map-related quantities of the selected WISDOM galaxies.}
 \begin{tabular}{lccccc}
  \hline
  Galaxy &  Beam size &  Pixel size & Sensitivity limit\\
  & (arcsec$^2$) & (arcsec) & ($\Msun$ pc$^{-2}$) \\
  \hline
FRL 0049 &  0.216 $\times$ 0.160  & 0.020 & 29.5 \\
NGC 0383 & 0.175 $\times$ 0.101  & 0.035 & 33.1 \\
NGC 0404 &  0.078 $\times$ 0.037 & 0.020  & 36.6 \\
NGC 0524 & 0.377 $\times$ 0.280  & 0.100  & 8.43 \\
NGC 1387 & 0.455 $\times$ 0.384  & 0.100  & 8.35 \\
NGC 1574 & 0.201 $\times$ 0.136 &  0.035  & 39.1 \\
NGC 4429 & 0.178 $\times$ 0.141  & 0.050 & 15.5 \\
NGC 4501 & 0.669 $\times$ 0.592 &  0.110  & 4.43 \\
NGC 4826 & 0.211 $\times$ 0.160 & 0.040 & 69.7 \\
NGC 5806 & 0.328 $\times$ 0.273 & 0.050  & 16.0 \\
NGC 6753 & 0.222 $\times$ 0.088 & 0.030  & 68.8 \\
NGC 6958 & 0.184 $\times$ 0.089 &  0.075 & 97.9 \\ 
  \hline
\multicolumn{4}{p{0.85\linewidth}}{\footnotesize{\textit{Notes:} For each galaxy, columns list the galaxy name, synthesised beam size (FWHM), pixel size and the molecular gas mass surface density sensitivity (i.e. root-mean-square noise of the observations).}}
 \end{tabular}
    \label{tab:obs_stats}
\end{table}

\begin{figure*}
    \centering
    \includegraphics[width=0.91\linewidth]{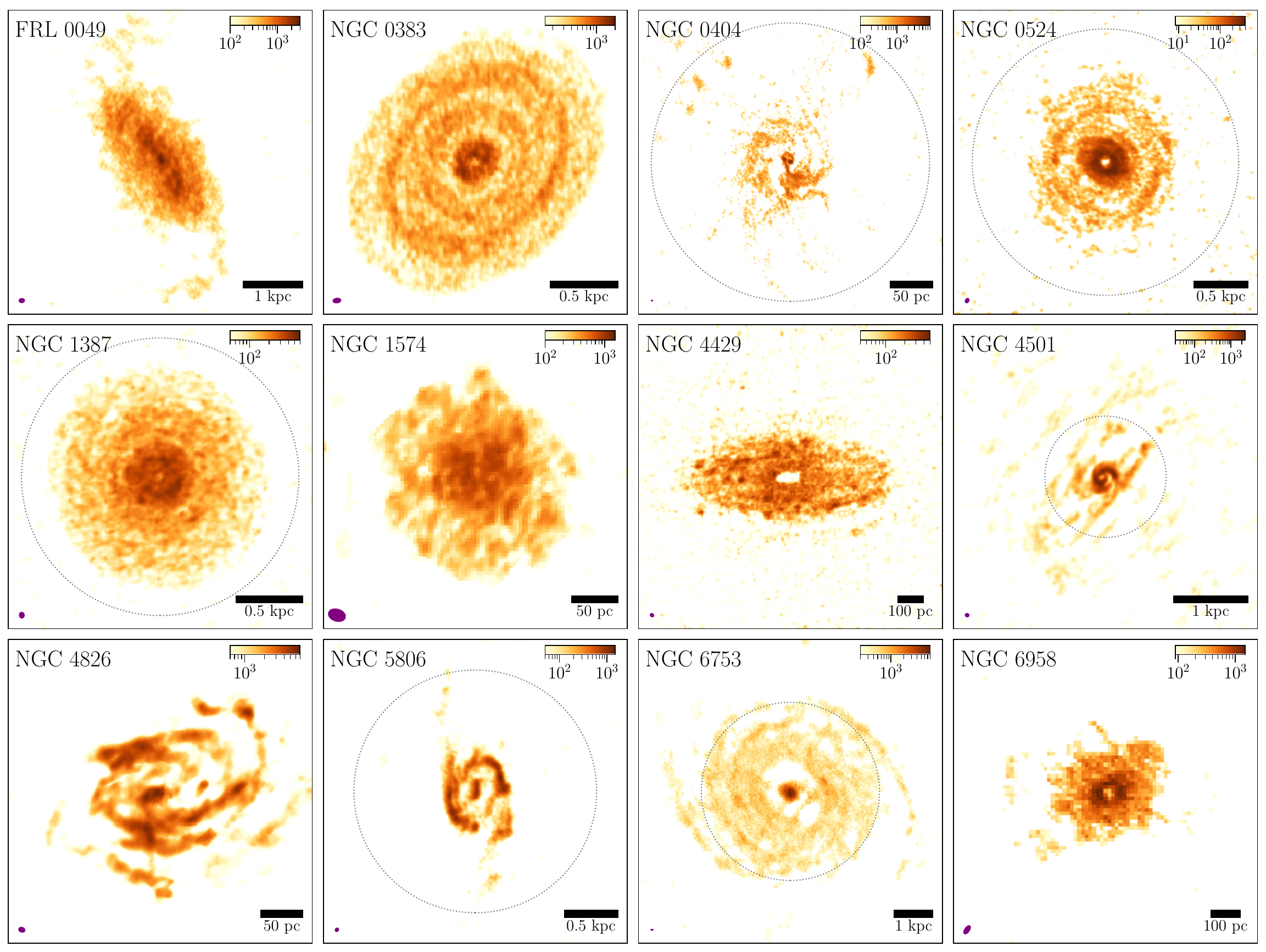}
    \caption{Moment zero maps of all WISDOM galaxies considered in this paper. In the top-right corner of each map we show the mass surface density scaling (conversion from original brightness units as described in Section~\ref{ss:Obs}). The scale bar in the bottom-right corner of each map gives and indication of the size of the visible circumnuclear gas disc. The purple ellipse in the bottom-left corner of each map shows the size of the synthesised beam. A black dashed circle indicates the $\ang{;;11}$ large scale limit of the power spectrum fit.}
    \label{fig:gasSD_obs}
\end{figure*}

We show the moment zero maps of all selected galaxies in Figure~\ref{fig:gasSD_obs}. Given that the primary aim of the WISDOM project is to dynamically measure SMBH masses in nearby galaxies, the resultant set of observed circumnuclear gas discs is quite heterogeneous. The molecular gas surface density sensitivity of the observations ranges from approximately 4 to 98 $\Msun ~ \pc^{-2}$, while the spatial resolution varies between approximately 1 to 80 pc. The extent and morphology of the circumnuclear gas reservoir also differ significantly across the twelve galaxies. 
As quantified with non-parametric morphological indicatiors by \citet{Davis2022}, the circumnuclear gas discs shown in Figure~\ref{fig:gasSD_obs} largely follow the prediction of \citet{Gensior2020}, that a bulge-dominated galaxy with high central stellar mass surface density will have a smooth central gas reservoir. To quantify this further, we hereafter turn to computing the mass surface density power spectra for all simulation snapshots and selected WISDOM galaxies. 

\subsection{Computing power spectra}\label{ss:PS}
The power spectra are computed using the python package \texttt{TurbuStat}\footnote{\url{https://turbustat.readthedocs.io/en/latest/index.html}} \citep{Koch2020}. For each map, the two-dimensional power spectrum is obtained by multiplying the Fourier transform of the mass surface density map by its complex conjugate. In this work, we focus on the one-dimensional (1D) power spectrum, which is generated by azimuthally averaging the two-dimensional power spectrum of each map. To simplify the comparison between simulations and observations, we fit a single power-law to each power spectrum:
\begin{equation}\label{eq:III_plaw}
    P(k) = A k^{-\beta},
\end{equation}
where $A$ is the amplitude of the power-law, $k$ the wavenumber, and $\beta$ the power-law (i.e. power spectrum) slope. As discussed in Section~\ref{s:Intro}, a broken power-law occasionally provides a better fit to the data \citep[e.g.][]{Dutta2009a,Grisdale2017}, and we will discuss individual deviations from a single power-law briefly in the Section~\ref{ss:PS_obs} and in more detail in Appendix~\ref{A:brokenpowerlaw}. 

Individual bright emission regions can dominate a power spectrum on small scales, to the extent that they can induce a bump and break in the power spectrum \citep[e.g.][]{Willet2005,Koch2020}. In our \emph{simulated} galaxy sample, we observe this behaviour in all of our bulge-dominated galaxies, due to the extremely dense region at the centre. There, gas accumulates continuously because star formation is suppressed and the simulations do not include AGN feedback, which could remove (some of) the innermost gas \citep{Gensior2020}. Therefore, we mask the centre of each simulated galaxy using an inverted cosine bell window function and refer the reader to Appendix~\ref{As:Cmask} for more details. To keep the analysis as consistent as possible, we similarly mask the centre of each WISDOM zeroth-moment map, where the extent of the central region masked depends on the size of the map. To avoid contamination of the power spectra by Gibbs ringing (stripes in the two-dimensional power spectrum; artefacts caused by a sharp cut-off of the emission), we additionally use a Tukey filter with $\alpha=0.1$ to taper the edges of simulated and observed gas maps. The $\alpha$ value of the Tukey filter tapering was determined based on a visual inspection of the two-dimensional power spectrum, where $\alpha=0.1$ is the most conservative value that fully removes the effects of Gibbs ringing. Changing the value of the Tukey filter $\alpha$ to a slightly larger value of 0.15 does not significantly affect the best-fitting power-law slopes. When varying $\alpha$, $\beta$ changes by $\lesssim$0.05, i.e. within the uncertainties on the power spectrum fit.

The fit to each power spectrum is limited by the beam size at small spatial scales and by the spatial scale within which all flux has been recovered at large scales. Specifically, here, we limit the fit to spatial scales larger than $3~\rm{FWHM_{\rm Gauss}}/\sqrt{8\log2}$, where $\rm{FWHM_{\rm Gauss}}$ is the FWHM of the Gaussian smoothing kernel used on the simulations, or the geometric average of the beam for the WISDOM observations which has not been circularised to maintain resolution. This means we avoid contamination of the power spectrum by beam effects\footnote{We have verified that including a beam response term in the power spectrum model and fitting within the same limits does not significantly affect the slope of the best-fitting power-law, compared to the fit to Equation~\ref{eq:III_plaw}.} and noise from pixelation \citep{Koch2020}. At large scales the extent of the fit is determined by the extent of the map (simulations and observations where the largest scale is smaller than $11''$), or by the spatial scale corresponding to $11''$. The exact fit range is stated in Table~\ref{tab:sims_pspecfit} for the simulations and in Table~\ref{tab:obs_pspecfit} for the observations.  

\section{Power spectra} \label{s:PS}

\subsection{Simulations} \label{ss:PS_sims}
\begin{figure*}
    \centering
    \includegraphics[width=.97\linewidth]{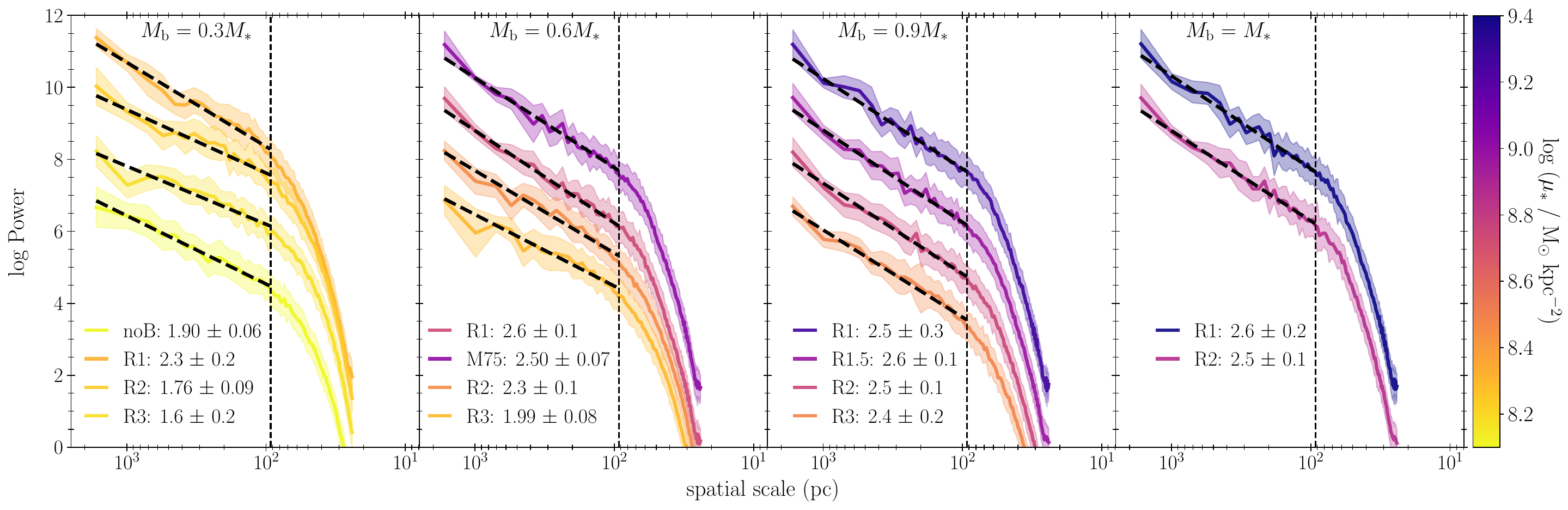}
    \caption{Gas mass surface density power spectra (solid lines colour-coded by the central stellar mass surface density of the simulated galaxy, with uncertainties as shading), computed from the snapshots 600 Myr after the start of the simulations. Each panel compares the effect of changing bulge scale radius (1 -- 3 kpc) at constant bulge mass, with panels from left to right showing simulations with bulges containing 30, 60, 90 and 100~per~cent of the initial stellar mass. The one bulge-less galaxy is shown with the leftmost panel and the single $\Mb=0.75\Mstar$ bulge is shown in the middle left panel. In each panel, the best-fitting power-laws are shown as black dashed lines, the vertical black dashed line indicates the smallest spatial scale considered for the fit, and the legend lists the best-fitting power-law slopes with their 1$\sigma$ uncertainties. All galaxies have a similar amount of power (arbitrary unit), thus individual power spectra are vertically offset from each other for visibility.}
    \label{fig:pspec_sims_3panel}
\end{figure*}

We first show example power spectra, with the power-law fits of Equation~\ref{eq:III_plaw} overlaid, for the suite of simulations at 600 Myr in Figure~\ref{fig:pspec_sims_3panel}. In each panel (of fixed initial bulge mass, increasing from the left to the right panel), the power spectra of different initial bulge scale radii are shown as solid lines, colour-coded by their central stellar mass surface densities. On each power spectrum the power-law fit is overlaid as a black dashed line. The shaded area around each solid line indicates the standard deviation of the azimuthal power in the two-dimensional power spectrum at this spatial scale. The vertical dashed line indicates the smallest spatial scale considered for the fit. As detailed in Section~\ref{ss:PS}, smaller scales are too affected by the beam to be included in the fit.

The panels in Figure~\ref{fig:pspec_sims_3panel} represent the different bulge mass fractions of 30 (left), 60 (centre left), 90 (centre right) and 100 (right) per~cent. All simulated galaxies are well fitted by a single power-law within the uncertainties of the azimuthal averages. In each panel, the power spectrum slopes increase from the more extended ($\Rb = 3~\kpc$) toward the more compact ($\Rb = 1~\kpc$) bulge. The slopes also increase from the left panel to right panel, demonstrating that a relatively more massive bulge (as the bulge-to-disc ratio also changes with increasing bulge mass) with the same scale radius results in a steeper power spectrum. This steepening of the power-law is expected if fragmentation is progressively suppressed \citep{Grisdale2017}. The trend shown in Figure~\ref{fig:pspec_sims_3panel} quantitatively confirms the qualitative impression from Figure~\ref{fig:gasSD_sims} and the results of the non-parametric morphological indicator study \citep{Davis2022} that galaxies with more compact and massive bulges, and thus with higher central stellar mass surface densities $\mus$, host smoother central gas reservoirs in which fragmentation is suppressed, compared to the disc-dominated galaxies that have clumpier and more sub-structured ISM. 

The power spectrum of the bulge-less galaxy is a mild outlier from the aforementioned trend, in that it has a slope that is steeper than that of two low-$\mus$, bulge-containing galaxy, even when accounting for the uncertainty of the fit. However, it is also the galaxy with the highest SFR of the simulated sample \citep{Gensior2020}, and therefore has the most SN feedback. This feedback contributes to the gas turbulence \citep[e.g.][]{Krumholz2018}, leading to a steeper slope \citep[see also e.g.][]{Walker2014}. 

Table~\ref{tab:sims_pspecfit} summarises the time-averaged results of the power spectrum fits to all simulated galaxies. The trend observed in Figure~\ref{fig:pspec_sims_3panel}, that galaxies with higher $\mus$ have steeper power spectra, also holds true when averaged across 700 Myr. In addition to the time-averaged best-fitting slope and its uncertainty, we also list the time-averaged uncertainty on the slope fit for each simulation.
Table~\ref{tab:sims_pspecfit} also confirms that the average power spectrum slope of the noB run agrees with that of the runs B\_M30\_R3, B\_M30\_R2 and B\_M60\_R3 within their variations over time and/or the uncertainties on the fits.

\begin{table}
\caption{Results of the power-law fits to the 1D power spectra of all simulations}
 \begin{tabular}{lcccc}
  \hline
  Simulation & $\left<\beta\right>$ & $\left<\rm{Fit}\:\rm{uncertainty}\right>$ &  Fit range  \\
   & & $1\sigma$ & (pc)\\
  \hline
noB & $1.84 \pm 0.04$ & 0.11 & 83 -- 1500 \\
B\_M30\_R1 & $2.32 \pm 0.06$ & 0.15 & 83 -- 1500 \\
B\_M30\_R2 & $1.78 \pm 0.06$ & 0.16 & 83 -- 1500 \\
B\_M30\_R3 & $1.77 \pm 0.04$ & 0.12 & 83 -- 1500 \\
B\_M60\_R1 & $2.63 \pm 0.04$ & 0.11 & 83 -- 1500 \\
B\_M60\_R2 & $2.29 \pm 0.05$ & 0.14 & 83 -- 1500 \\
B\_M60\_R3 & $1.88 \pm 0.04$ & 0.12 & 83 -- 1500 \\
B\_M75\_R1 & $2.50 \pm 0.06$ & 0.15 & 83 -- 1500 \\
B\_M90\_R1 & $2.49 \pm 0.09$ & 0.26 & 83 -- 1500 \\
B\_M90\_R1.5 & $2.54 \pm 0.05$ & 0.15 & 83 -- 1500 \\
B\_M90\_R2 & $2.51 \pm 0.06$ & 0.16 & 83 -- 1500 \\
B\_M90\_R3 & $2.43 \pm 0.06$ & 0.16 & 83 -- 1500 \\
B\_M100\_R1 & $2.57 \pm 0.06$ & 0.16 & 83 -- 1500 \\
B\_M100\_R2 & $2.52 \pm 0.05$ & 0.14 & 83 -- 1500 \\

  \hline
  \multicolumn{4}{p{\linewidth}}{\footnotesize{\textit{Notes:} For each simulation, column 1 lists the name, column 2 lists the time-averaged best-fitting slope and its uncertainty, column 3 lists the time-averaged uncertainties on the slope listed in column 2, and column 4 lists the spatial scales across which a power-law is fitted to the 1D power spectrum.}}
 \end{tabular}
 \label{tab:sims_pspecfit}
\end{table}

\begin{figure}
    \centering
    \includegraphics[trim={0cm 0.3cm 0cm 0cm}, clip=true,width=1.\linewidth]{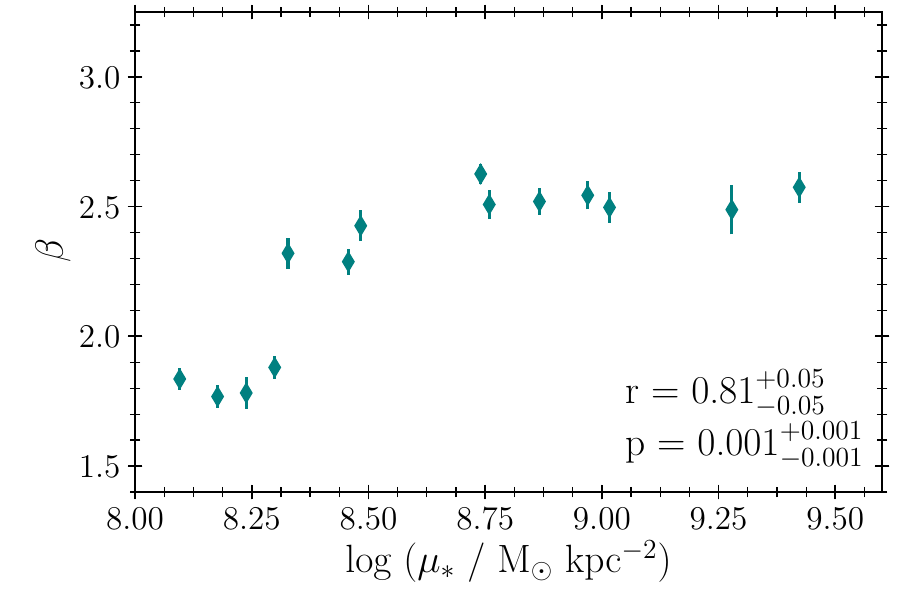}
    \caption{Time-averaged slopes ($\beta$) of the power-laws best-fitting the power spectra of the simulated galaxies as a function of the galaxies' central stellar mass surface densities ($\mus$). Error bars indicate the uncertainties on the time-averaged power-law slopes. As indicated by the Spearman rank correlation coefficient and p-value listed in the bottom-right corner, there is a statistically-significant correlation between the best-fitting slope of the power spectrum and the central stellar mass surface density of a galaxy.}
    \label{fig:pspec_slopes_mus_sims}
\end{figure}

To be more quantitative in linking the galactic gravitational potential to the power spectrum, we plot the power spectrum slopes as a function of the central stellar mass surface densities of all simulated galaxies (calculated as described in Section~\ref{ss:Sims}) in Figure~\ref{fig:pspec_slopes_mus_sims}. 
There is a clear trend of increasing $\beta$ with increasing $\mus$ for $\log (\mus/\Msun~\kpc^{-2}) \leq 8.75$, which flattens at $\log (\mus/\Msun~\kpc^{-2}) > 8.75$. However, it is unclear whether this flattening is physical in origin, or whether it is due to the limited spatial resolution of the simulations, or is caused by other parameter choices. While the power-law fit is restricted to spatial scales larger than the minimum gravitational softening, the sub-structure on smaller scales might be affected by the gravitational softening choice. Small scale fragmentation might be even more suppressed in the highest $\mus$ galaxies which would lead to a steeper power spectrum, and thus a continuous trend of increasing $\beta$ with $\mus$, however higher resolution simulations would be required to test this. Interestingly, the flattening of the $\beta$-$\mus$ trend occurs at a slope of $\sim$2.6, which is close to that expected of two-dimensional \citet{Kolmogorov1941} turbulence (8/3). If physical, this flattening could therefore indicate that high shear eventually drives incompressible turbulence. We use Spearman's rank correlation coefficient to assess the strength and statistical significance of this correlation. To take into account the uncertainties on the time-averaged slopes, we perform 1000 Monte-Carlo simulations assuming that the power spectrum slope of each data point is well described by a Gaussian distribution. This allows us to quote both Spearman's rank correlation coefficient and p-value with uncertainties, as given by the average and the 16th-to-84th percentile of the Monte-Carlo simulation. A Spearman rank correlation coefficient of $r = 0.81_{-0.05}^{+0.05}$ with $p = 0.001_{-0.001}^{+0.001}$ confirms that the correlation between $\beta$ and central stellar mass surface density is strong and statistically significant. This directly links the galactic gravitational potential to the suppression of fragmentation via the power spectrum slope. It also confirms that shear from a high-$\mus$ potential does not only drive turbulence that suppresses star formation, but it also tears apart clouds and suppresses fragmentation. This is a strong prediction from the simulations, to which we can compare the WISDOM observations. 

\subsection{Observations} \label{ss:PS_obs}
\begin{figure*}
    \centering
    \includegraphics
    [width=.97\linewidth]{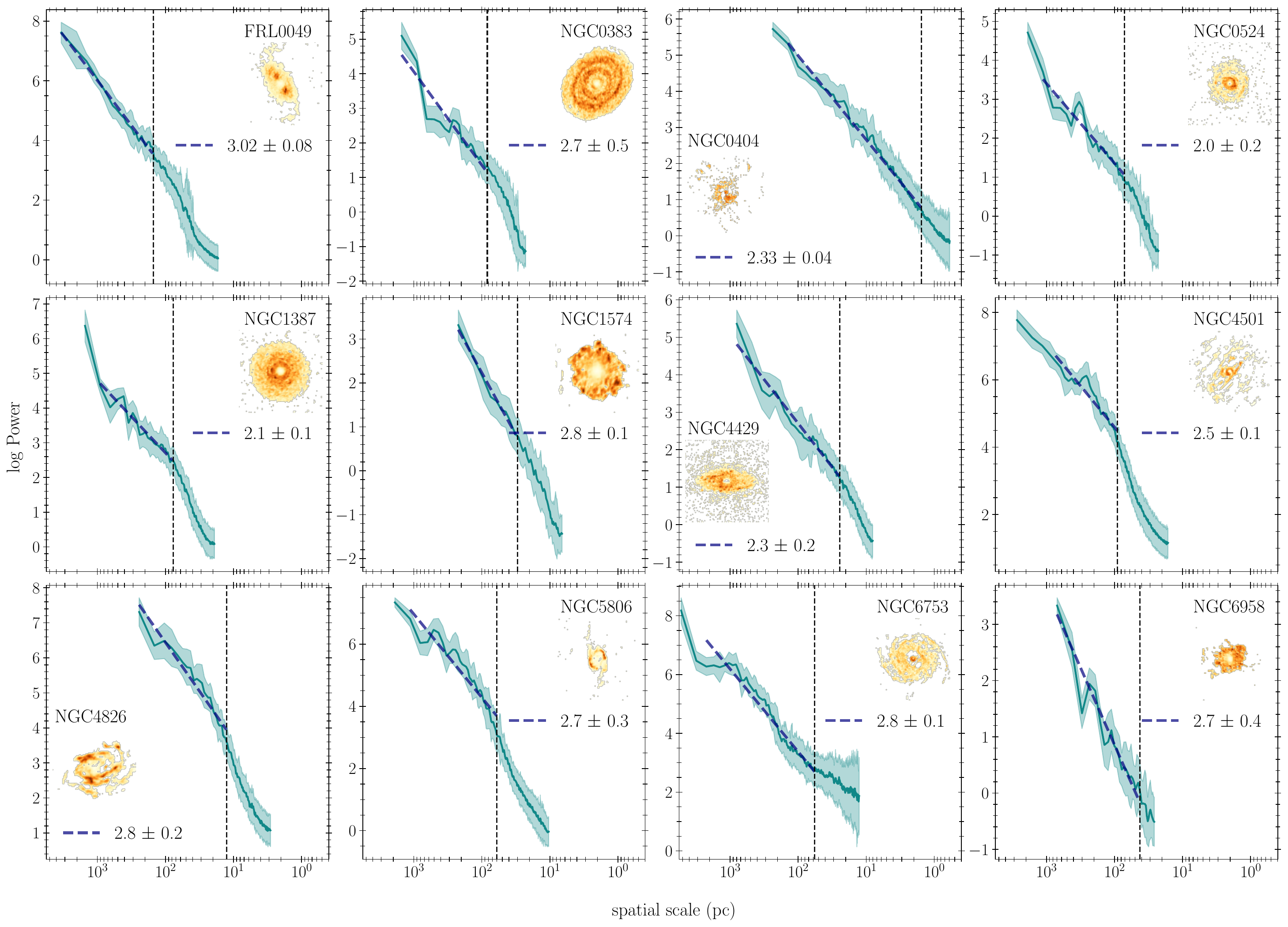}
    \caption{Gas mass surface density power spectra of the WISDOM galaxies considered in this paper (solid teal lines, with uncertainties as shading), plotted with unique spatial scale for ease of comparison. In each panel, the best-fitting power-law is shown as navy dashed line, the vertical black dashed line indicates the smallest spatial scale considered for the fit and the legend lists the best-fitting power-law slope with its 1$\sigma$ uncertainty. The inset shows the centre-masked moment zero map of the galaxy, from which the power spectrum was computed.}
    \label{fig:pspec_obs_indv}
\end{figure*}

Figure~\ref{fig:pspec_obs_indv} shows the power spectra of the WISDOM galaxies considered in this paper. All spectra are plotted on the same spatial scale, both to make comparison easier and to highlight the different spatial scales under consideration (that are function of the galaxy distance, map size and synthesised beam). In each panel, the navy dashed line shows the best-fitting power-law to the power spectrum, an inset shows the centre-masked zeroth-moment map from which the power spectrum was computed and the black vertical dashed line indicates the smallest spatial scale considered for the fit. We summarise the results of the 1D power spectrum fits of the WISDOM galaxies in Table~\ref{tab:obs_pspecfit}.

\begin{table}
\centering
\caption{Results of the power-law fits to the 1D power spectra of all selected WISDOM galaxies}
 \begin{tabular}{ccc}
  \hline
  Galaxy & $\beta$ & Fit range\\
   & & (pc) \\
  \hline
FRL 0049 & $3.02 \pm 0.08$ & 151 -- 3450 \\
NGC 0383 & $2.70 \pm 0.45$ & 83 -- 1518 \\
NGC 0404 & $2.33 \pm 0.04$ & 2 -- 163 \\
NGC 0524 & $2.05 \pm 0.18$ & 71 -- 1243 \\
NGC 1387 & $2.09 \pm 0.14$ & 77 -- 1045 \\
NGC 1574 & $2.83 \pm 0.12$ & 30 -- 222 \\
NGC 4429 & $2.34 \pm 0.22$ & 24 -- 809 \\
NGC 4501 & $2.45 \pm 0.15$ & 90 -- 816 \\
NGC 4826 & $2.78 \pm 0.23$ & 13 -- 246 \\
NGC 5806 & $2.69 \pm 0.29$ & 60 -- 1141 \\
NGC 6753 & $2.81 \pm 0.13$ & 57 -- 2325 \\
NGC 6958 & $2.75 \pm 0.44$ & 42 -- 700 \\
  \hline
  \multicolumn{3}{p{0.75\linewidth}}{\footnotesize{\textit{Notes:} For each galaxy, column 1 lists the name, column 2 lists the best-fitting slope and the $1\sigma$ uncertainty on the fit. Column 3 lists the spatial scales across which a power-law was fitted to the 1D power spectrum.}} 
 \end{tabular}
 \label{tab:obs_pspecfit}
\end{table}

Only half of the galaxy power spectra (FRL 0049, NGC 0404, NGC 1387, NGC 1574, NGC 4429 and NGC 4826) are well described by a single power-law. The other spectra show very pronounced bumps and/or breaks. In NGC 0383, NGC 0524, NGC 4501, NGC 5806, and NGC 6958 this is likely caused by ring structures.
NGC 6753 possesses a very bright central region within a very low density region, itself surrounded by a low-density gas disc with weak spiral arms. These likely give rise to the complicated shape of the power spectrum, with multiple break points. 

To assess the validity of our single power-law fits, we recompute the fit to the power spectra of the observations, allowing for a break, i.e. a dual power-law fit using the \texttt{turbustat} fitting routine and a segmented linear fit. Comparing the Bayesian information criterion of the fits with and without a break yields only three galaxies, NGC 4501, NGC 4826 and NGC 5806, whose power spectra are better fit by broken power-laws. A break in a surface mass density power spectrum is thought to indicate the scale height of the gas disc, as turbulence transitions from three-dimensional turbulence on small scales to two-dimensional turbulence on large scales \citep[e.g.][]{Dutta2009b}. The break scale of the NGC 4826 power-law is at 27 pc, which would imply an extremely thin molecular gas disc. \citet{Yim2014} measured molecular gas scale heights of $31\pm8$ and $33\pm8$ pc for two other galaxies, NGC 5907 and NGC 4565, respectively, but CO scale heights of 100--200 pc are more common \citep{Wilson2019}.  However, 27 pc approximately matches the size of the dense clouds seen in the zeroth-moment map, which may cause the break \citep{Koertgen2021}. The breaks in NGC 4501 and NGC 5806 occur at 239 pc and 191 pc, respectively, and could more plausibly be linked to the scale height of their molecular gas discs. Alternatively, the fits might be affected by the prominent ring structures. However, even the best-fitting single power-laws tend to be within the uncertainties associated with the 1D gas mass surface density power spectrum of these galaxies. Therefore, we use the best-fitting slope of the single power-law for all observed galaxies for consistency and we refer the reader to Appendix~\ref{A:brokenpowerlaw} for an in-depth discussion of the multi-component power-law fits.

\begin{figure}
    \centering
    \includegraphics[trim={0cm 0.3cm 0cm 0cm}, clip=true,width=1.\linewidth]{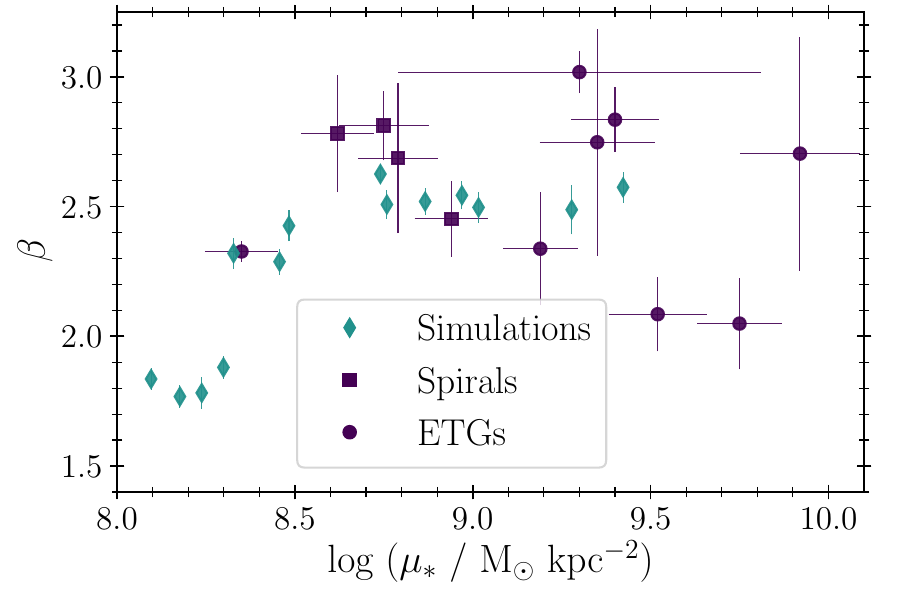}
    \caption{Slopes ($\beta$) of the power-laws best-fitting the power spectra of the simulated (teal) and WISDOM (purple) galaxies as a function of their central stellar mass surface densities ($\mus$). The vertical error bars indicate the uncertainties on the time-averaged best-fitting power-law slopes (simulations) or the 1$\sigma$ uncertainties on the best-fitting power-law slopes (observations).}
    \label{fig:pspec_slopes_mus_obs_sims}
\end{figure}

In Figure~\ref{fig:pspec_slopes_mus_obs_sims} we overplot the WISDOM galaxies' best-fitting power spectrum slopes as a function of their central stellar mass surface densities on the simulation data. Contrary to the clear trend present in the simulations, the slopes of the observed galaxies have more scatter with respect to $\mus$. This is a clear departure from the prediction of the simulations. There is neither a trend of power spectrum slope with central stellar mass surface density for the whole dataset of WISDOM galaxies ($r = -0.14_{-0.24}^{+0.24}$, $p = 0.55_{-0.34}^{+0.31}$), nor for the sub sets of spiral galaxies ($r = 0.02_{-0.52}^{+0.48}$, $p = 0.57_{-0.39}^{+0.30}$) or ETGs ($r = -0.09_{-0.31}^{+0.33}$, $p = 0.57_{-0.31}^{+0.29}$). The lack of a correlation with $\mus$ could in part stem from the unsuitability of some power spectra to being fit by a single power-law. Alternatively, the observations might follow the trend of the simulations (at $\log (\mus/\Msun~\kpc^{-2}) > 8.75$), but the measurement uncertainties on $\beta$ and $\mus$ as well as potential scatter from physical effects not included in the simulations could obscure it. We will discuss the discrepancies between power spectra of the simulations and WISDOM galaxies in more depth in Section~\ref{ss:sims_vs_obs_diff}. However, before doing so, we wish to investigate whether the power spectrum slopes of the WISDOM galaxies are correlated with other quantities. 

\section{Discussion} \label{s:Disc}

\subsection{Correlations with the slopes of the WISDOM power spectra?} \label{ss:obs_pspec_cors}
In an attempt to identify the driver of turbulence in the WISDOM galaxies, we have checked for correlations between the power spectrum slope and a large number of quantities. Some are observational (beam size, extent of the fit, lower spatial scale limit of the fit, sensitivity of observations) while others are galaxy properties (gas mass, mean central molecular gas mass surface density, stellar mass, gas fraction central gas-to-stellar mass surface density ratio, SFR, specific SFR, star formation efficiency, Gini, Asymmetry, Smoothness and the stellar-to-gas \citet{Toomre1964} Q ratio). Table~\ref{tab:obs_cor_all} lists all Spearman rank correlation coefficients and corresponding p-values. If we define $p\leq 0.1$ as being statistically significant, we find a single statistically-significant correlation (with mean central molecular gas mass surface density) for the observations. This statistically-significant correlation, as well as some other quantities of interest, are discussed in more detail below. We focus on the observations here, because the simulations only differ in the underlying stellar potential by construction. They are initialised with the same gas fraction and the same radial gas extent, all of which only evolve very slightly, thus resulting in a very narrow range of mean central molecular gas mass surface densities \citep[see also Figure 4 in][]{Davis2022}. Differences in the simulated galaxies' (s)SFR are driven by changes in $\mus$ (see Figure 15 in \citealt{Gensior2020}), therefore any trend of the physical quantities investigated for the observations, if present in the simulations, will be a result of the correlation with the central stellar mass surface density discussed in Section~\ref{ss:PS_sims}.

\begin{figure*}
    \centering
    \includegraphics[trim={2.2cm 1cm 0.5cm 0.5cm}, clip=true ,width=0.95\linewidth]{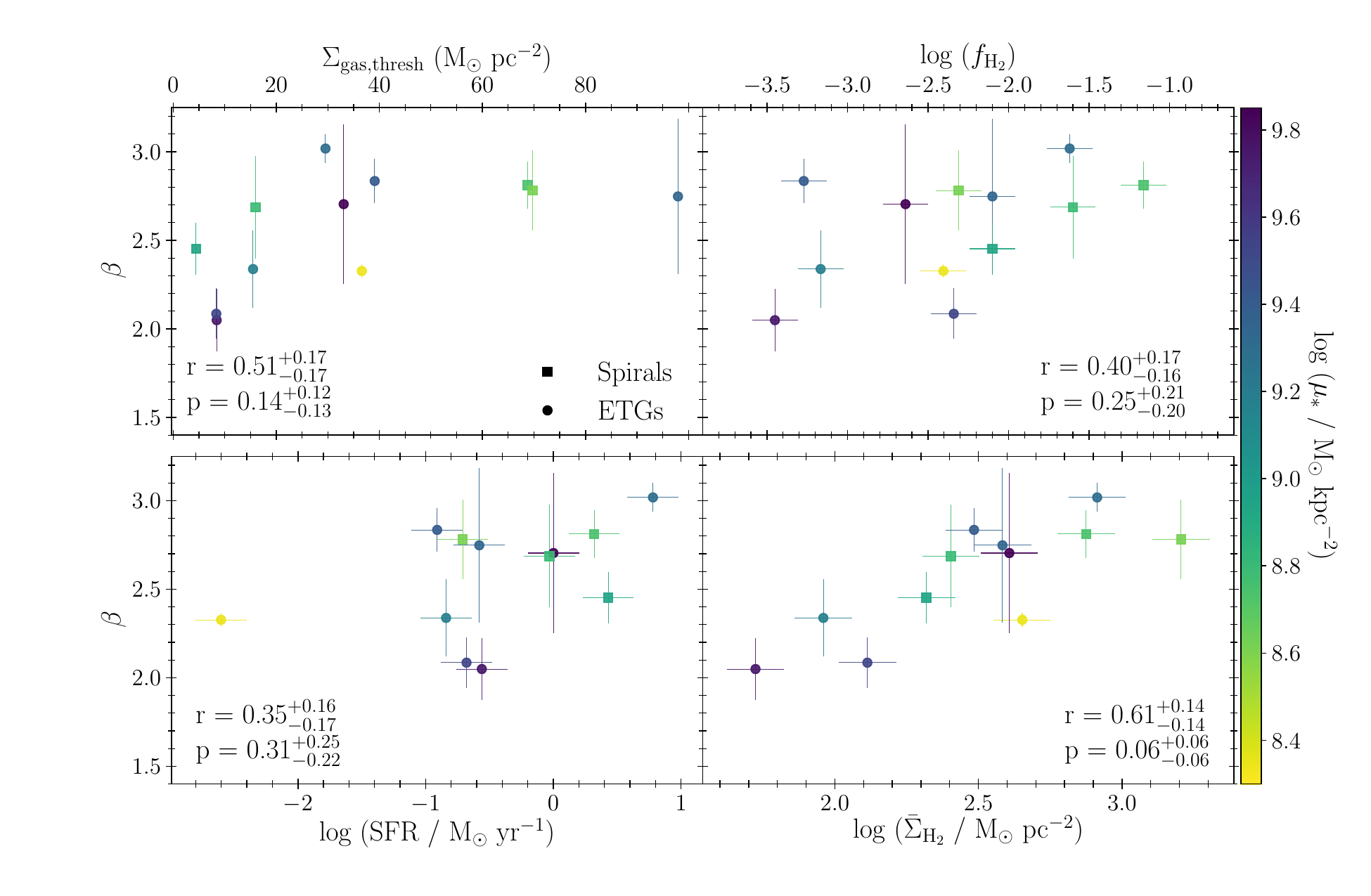}
    \caption{Slopes ($\beta$) of the power-laws best-fitting the power spectra of the WISDOM galaxies as a function of the galaxies' mass surface density sensitivities ($\Sigma_{\rm gas, thresh}$, top left), molecular gas fraction ($f_{\rm H_2}$, top right), SFR (bottom left) and average molecular gas mass surface density within the central kpc ($\bar{\Sigma}_{\rm H_2}$, bottom right). The data points are colour-coded by their central stellar mass surface densities ($\mus$). Vertical error bars indicate the 1$\sigma$ uncertainties on the best-fitting power-law slopes and horizontal error bars show the uncertainty on the respective quantity. The Spearman rank correlation coefficient and p-value for each correlation are shown in each panel. There only statistically significant correlation is that between power spectrum slope and the average central molecular gas mass surface density of the observations.}
    \label{fig:pspec_obs_4panel_muscol}
\end{figure*}

\subsubsection{Sensitivity of the observations}\label{sss:sens}
\citet{Koch2020} highlighted that observational effects can influence a power spectrum, even on spatial scales larger than the nominal spatial resolution (geometric beam size; see also \citealt{Koertgen2021}). With sensitivities ($\Sigma_{\rm gas, thresh}$) that vary by more than an order of magnitude across WISDOM galaxies, from 4 (NGC 4501) to 97 $\Msun~\pc^{-2}$ (NGC 6958), it is worth examining whether the sensitivities of the WISDOM galaxies affect the derived power spectra. The top left panel of Figure~\ref{fig:pspec_obs_4panel_muscol} reveals a putative trend between the molecular gas mass surface density sensitivity of the dataset and the best-fitting power-law slope of the resultant power spectrum. The power spectrum appears to steepen with worsening (i.e. higher surface density) sensitivity. 
However, the $\beta - \Sigma_{\rm gas, thresh}$ trend has a Spearman rank correlation coefficient $r = 0.52_{-0.17}^{+0.16}$ and p-value $p = 0.14_{-0.13}^{+0.14}$, so it is not statistically significant. To nevertheless test if any of our results are affected by this possible bias, we have performed the observational analysis again with all zeroth-moment maps clipped to a molecular gas mass surface density of 70 $\Msun~\pc^{-2}$, to mimic a uniform sensitivity. It turns out even this crude approximation only has a negligible effect on the correlation between best-fitting power-law slope and central stellar mass surface density shown in Figure~\ref{fig:pspec_slopes_mus_obs_sims}. Despite this, we advise to be mindful of potential effects when comparing the power spectra of a heterogeneous set of observations. 

\subsubsection{Star formation rate}\label{sss:SFR}
Star formation is one of the primary drivers of feedback and thus turbulence \citep[e.g.][]{MacLow2004,Hennebelle2012}. Here we examine whether there is a correlation between the power spectrum slopes and the SFRs of the selected WISDOM galaxies. With our sample comprising ETGs and spirals, we cover several decades in SFR, thus expect to have sufficient dynamic range to identify a correlation, should it exist. However, the bottom left panel of Figure~\ref{fig:pspec_obs_4panel_muscol} shows that there is no trend between the power spectrum slope and the SFR of a galaxy. This is confirmed by a Spearman rank correlation coefficient of $r = 0.35_{-0.16}^{+0.17}$ with $p = 0.31_{-0.23}^{+0.25}$. 

\subsubsection{Gas fraction}\label{sss:gas_frac}
Dynamical suppression has a strong dependence on the gas fractions of galaxies \citep{Martig2013,Gensior2021}. Additionally, \citet{Gensior2021} demonstrated that the morphology of the circumnuclear gas reservoir can be strongly affected by the gas-to-stellar mass ratio of the galaxy. At fixed (high) central stellar mass surface density, the smooth circumnuclear region of the gas disc decreases in extent until finally the entire ISM is porous and sub-structured for increasing initial cold gas-to-stellar mass ratio from 1 to 20~per~cent. For simplicity we have restricted ourselves in this study to the analysis of the \citet{Gensior2020} simulations with a constant initial cold gas-to-stellar mass ratio of 0.05, but the observed galaxies have a range of gas fractions. Therefore, we investigate the molecular gas-to-stellar mass ratios of the WISDOM galaxies in relation to their power spectrum slopes in the top right of Figure~\ref{fig:pspec_obs_4panel_muscol}. However, as evidenced by the Spearman rank correlation coefficient $r = 0.40_{-0.16}^{+0.17}$ with a $p = 0.25_{-0.20}^{+0.21}$, there is no statistically-significant trend. There is also no secondary trend with $\mus$, in contrast to the prediction from the simulations. 

\subsubsection{Mean central molecular gas mass surface density}\label{sss:mug}
The bottom right panel of Figure~\ref{fig:pspec_obs_4panel_muscol} shows the power spectrum slopes of the WISDOM galaxies as a function of their central mean molecular gas mass surface density, $\mug$. Following \citet{Davis2022}, $\mug$ is the average molecular gas mass surface density within an ellipse of semi-major axis 1 kpc, that is centred on the centre of the galaxy and oriented according to the galaxy's position angle. There is a trend of steeper power spectra with increasing mean central molecular gas mass surface density with a Spearman rank correlation coefficient $r = 0.61_{-0.13}^{+0.13}$ and $p = 0.06_{-0.05}^{+0.05}$. Therefore, this trend is statistically significant for our definition of $p < 0.1$. 
\citet{Davis2022} found similar correlations between the non-parametric morphological parameters and $\mug$, in that galaxies with higher central gas surface density had a smoother, less sub-structured ISM. A correlation with $\mug$ suggests that the self-gravity of the gas could play an important role in setting the structure of (and driving turbulence in) the ISM. However, naively one would expect the strong self-gravity to affect the ISM structure in the opposite way, through fragmentation, from which one would expect an anti-correlation between $\beta$ and $\mug$. Therefore, it is unclear whether self-gravity is the dominant driver of turbulence in these galaxies. \citet{Davis2022} analysed the stellar-to-gas \citet{Toomre1964} Q ratio to determine whether the stability of the system is dominated by the stellar or the gaseous component. They found that in $\approx$75~per~cent of galaxies the stability of the gas disc is dominated by the stellar component, suggesting that the $\mug$ trend is a secondary correlation with the gravitational potential. However, we find no trend between the power spectrum slopes and the stellar-to-gas Q ratios of the galaxies, nor with any other more direct tracers of the potential, implying that while the mean central gas mass surface density might weakly depend on the potential, it seems to be the more relevant quantity for the power spectrum.
Another possibility is that the $\beta$ correlates with $\mug$, as a tracer for the turbulent energy per unit area of gas, which depends on the gas surface density \citep[e.g.][]{Krumholz2018}. This potentially hints at gravitational momentum transport or accretion driven turbulence as the drivers of turbulence in the observed galaxies, explaining the correlation with $\mug$. 
However, the differences in the sensitivities of the observations (see Table~\ref{tab:obs_stats} and Section~\ref{sss:sens}) may affect the positive correlation between power spectrum slope and mean central molecular gas mass surface density. Therefore, a large(r), more homogeneous sample of observations is required to confirm this trend. The simulations do not show any trend with $\mug$ due to the small dynamic range of mean central gas mass surface densities they have by construction.

All data points in Figure~\ref{fig:pspec_obs_4panel_muscol} are colour-coded by their central stellar mass surface density, to see if any secondary correlations with $\mus$ exist. However, contrary to expectations from the simulations and the results of \citet{Davis2022}, no such secondary correlations exist, implying that the stellar potential has a smaller effect on the power spectra (and possibly driving turbulence) of the observed galaxies, than on the ISM structure as quantified through Gini, Asymmetry and Smoothness.  
\citet{Henshaw2020} found that analysis of the gas velocity structure, in addition to the gas density, is required to fully understand the density structure in different galactic environments. Therefore, the velocity spectra of the WISDOM galaxies could provide more insight into the power spectrum slope discrepancies of galaxies with similar discs (as per non-parametric morphology indicators, see \citealt{Davis2022}) and galactic properties - this will be pursued in future works.

\subsection{Differences between observations and simulations} \label{ss:sims_vs_obs_diff}
Figure~\ref{fig:pspec_slopes_mus_obs_sims} shows the power spectrum slopes of both simulated and real galaxies as a function of their central stellar mass surface densities. Shear induced by the deep gravitational potentials is the main driver of turbulence in the simulations, as indicated by the trend of steeper power spectra as a function of $\mus$ for these. The observations with central stellar mass surface densities in the range $8.3\leq \log(\mus/\Msun~\kpc^{-2}) \leq 8.8$ follow the simulated data. However, as pointed out in Section~\ref{ss:PS_obs}, instead of even steeper power spectra or a flattening of the trend for WISDOM galaxies with larger central stellar mass surface densities, the other eight galaxies are scattered in $\beta-\mus$ space. Nonetheless, they remain broadly consistent with the simulations. This could indicate that the flattening of $\beta$ to $\sim2.6$ at $\log(\mus/\Msun~\kpc^{-2}) \geq 8.75$ for the simulated galaxies is physical in nature, and in that case, given the close correspondence with the slope of two-dimensional \citet{Kolmogorov1941} turbulence, that high shear might eventually drive incompressible turbulence. In the remainder of this subsection, we will further discuss the differences between the simulations and observations, that may contribute to a more nuanced scenario and explain the larger scatter of the observations.

It should be kept in mind that the simulations are idealised galaxies in isolated boxes, where the only difference between the individual simulated galaxies is the distribution of the stellar matter. Therefore, their only sources of turbulence are shear, (self-)gravity and SN feedback. The simulated galaxies have not evolved over a Hubble time in a cosmological setting, i.e. they have never accreted gas from the cosmic web, undergone mergers or interactions with other galaxies, and by construction do not contain black holes. In contrast, each of the WISDOM galaxies hosts a SMBH, which is active in some of them (e.g. FRL 0049, NGC 0383 and NGC 6753). 
To our knowledge, the impact of an active galactic nucleus on the power spectrum of a galaxy has not yet been investigated. The radius of the SMBH sphere of influence tends to be comparable to the lower spatial-scale limit of the power spectrum fit, so the SMBHs direct influence on the potential should have little impact on the fit. However, it remains difficult to estimate the effect of an active SMBH on the power spectrum, as its feedback could drive turbulence in the gas \citep[e.g.][]{Venturi2021,Tamhane2022}. 

Cluster membership and related effects on galaxies can also affect their power spectra. Investigating the power spectrum of the Virgo cluster galaxy NGC 4254, \citet{Dutta2010} found that the power spectrum slope is different in the inner and outer regions of the galaxy (i.e. for different velocity channels), as well as in the velocity-integrated power spectrum. \citet{Dutta2010} attributed this difference to galaxy harassment. As most of the WISDOM galaxies considered here are members of a galaxy group or cluster, this could contribute to the scatter of their power spectrum slopes and the poor quality of the fit for some. It is also important to capture the effects of galaxy interactions and cosmological evolution in simulations, as demonstrated by comparing the the cold gas mass surface density power spectrum slope between observation and simulation of the Small Magellanic Cloud (SMC): With a (large-scale) slope of 1.7, \citet{Grisdale2017} found a much shallower power-law for their simulated SMC than that of \citet{Stanimirovic1999}, who measured a slope of 3.04 from \hi observations. \citet{Grisdale2017} argued that this discrepancy is the result of running an isolated galaxy simulation lacking tidal effects from galaxy interactions. Another potentially important source of turbulence is gas accretion both on the larger spatial scale for accretion onto the gas disc of the galaxy, but also on the smaller spatial scale accretion of gas onto molecular clouds \citep[e.g.][]{Klessen2010}, which also impacts the power spectrum slope \citep{Koertgen2021}.

In short, while there is broad agreement between the simulations and the WISDOM observations, the discrepancies are likely the result of several factors. In particular, the heterogeneity of the observations (e.g. sensitivity, see Section~\ref{ss:obs_pspec_cors}) could dilute the effect of the stellar potentials on the power spectrum slopes, and there are many effects not captured by the simulations (e.g. AGN feedback). A larger, more homogeneous sample of observations should be able to address some of these observational issues. Similarly, high-resolution cosmological zoom-in simulations should produce more realistic comparisons. 

\subsection{Comparison with the literature}
\subsubsection{Simulations}
Over the past decade, power spectra have been used to assess the necessity for and quality of feedback models in simulations. Most of these simulated galaxies have breaks in their power spectra at scales of several tens of pc to 1 kpc (e.g. \citealt{Bournaud2010,Combes2012,Renaud2013,Walker2014,Grisdale2017}; although the power spectrum break in \citealt{Renaud2013} is at a much smaller scale of about 1 pc). Unlike these, the power spectra of our simulations are well described by single power-laws, comparable to the two runs with less energetic feedback of \citet{Walker2014}. The slopes (at large scales in the case of broken power-law fits) range from 0.67 \citep{Renaud2013} to 4.5 \citep{Pilkington2011}, although they are mostly in the range 1.3 -- 2.6 \citep{Bournaud2010,Combes2012,Walker2014,Grisdale2017} for runs with stellar feedback. The latter range is in good agreement with that of our own simulated power spectra (1.8--2.7). A common trend in the aforementioned simulations is a steepening of the power spectra in simulations with (increasing) feedback. This may seem counter-intuitive when compared to our result of steeper power spectra with increasing suppression of fragmentation (and therefore suppression of star formation and subsequent feedback). However, there is a simple explanation for this: both strong feedback and shear from the galactic gravitational potential have the same effect. Shear-driven turbulence shifts the power in the gas from small to large scales by suppressing fragmentation and the associated formation of GMCs. Similarly, strong feedback shifts the power from small-scale to large scale by destroying clouds more effectively \citep[e.g.][]{Pilkington2011,Walker2014}. Some evidence of the effects of feedback are also present in our simulated galaxies. Indeed, Table~\ref{tab:sims_pspecfit} shows that the simulated bulge-less galaxy has a steeper power spectrum than a couple of other simulated low-$\mus$ galaxies, a direct consequence of its higher SFR. 

\subsubsection{Observations}
A variety of tracers have been used in studies of extragalactic power spectra. However, only M~31 \citep{Koch2020} and M~33 \citep{Combes2012,Koch2020} have been previously analysed in CO. With slopes of 1.59 for M31 and 1.5 (\citealt{Combes2012}) or 0.91 (\citealt{Koch2020}) for M~33, these power spectra are much shallower than those measured here for the WISDOM galaxies. If ISM turbulence is truly scale-free, then it should not matter that the length scales considered are vastly different. The M~31 and M~33 data probe scales ranging from hundreds of pc to several tens of kpc, whereas the circumnuclear gas discs of some of our galaxies only span a few hundred parsecs, and even in the most extreme case the largest length scale probed is 3.4 kpc. However, like \citet{Koch2020}, we find that the morphology of the molecular gas disc is important to determine its power spectrum slope and shape. With often only a small disc detected, it is difficult to compare\footnote{We see evidence from the simulations that considering larger regions in the power spectrum fits leads to a flattening of the power spectra, as more of the clumpy, sub-structured ISM are included.}. Encouragingly \citet{Miville-Deschenes2010}, and \citet{Pingel2018} found power spectrum slopes of 2.7 when using dust and CO emission to study the Polaris flare and Perseus molecular cloud regions within the Milky Way, in better agreement with the slopes of the WISDOM galaxies than those of \citet{Combes2012} and \citet{Koch2020}.

The power spectrum slopes of our galaxies are more similar to those often measured from \hi, which tend to be in the range 1.5 -- 3.0 \citep[e.g.][and references therein]{Dutta2009a,Zhang2012,Dutta2013,Walker2014,Grisdale2017, Koch2020}. This raises an interesting question, namely whether the \hi power spectra of the WISDOM galaxies would be steeper or shallower than their CO counterparts. Turbulence theory predicts that the power spectrum of the molecular gas should have a steeper slope than that of its atomic counterpart \citep{Romeo2010}, although for M31 and M33 \citet{Koch2020} found that the \hi spectra are steeper. With the extremes of the \hi power spectrum slopes at 0.3 (\citealt{Dutta2013}) and 4.3 (\citealt{Zhang2012}), both steeper and shallower \hi power spectra for the WISDOM galaxies would not be without precedent\footnote{Having said that, only 10~per~cent of ETGs in cluster environments have \hi \citep[e.g.][]{Serra2012}, so measuring \hi power spectra is likely to be possible only for ETGs in lower density environments.}.

We do not have evidence of ubiquitous breaks in the power spectra, that would indicate transitions from two-dimensional turbulence in the planes of the galaxies to three-dimensional turbulence at the scale heights of the galaxies \citep[e.g.][]{Dutta2009b}. Instead, morphological features specific to each galaxy (such as rings) show up as bumps or wiggles in the power spectrum. Only three galaxies are better described by broken power-laws (than by single power-laws). For NGC 4826 the break scale is at roughly the size of the brightest clumps in the gas disc, making it more likely that the break in the power spectrum indicates a transition from turbulence in the large-scale ISM to that in the largest fragments, in good agreement with the findings of \citet{Koertgen2021}. The break scale in NGC 4501 and NGC 5806 may be at around around the disc scale height, but it might also be caused by the effect of the dominant ring on the power spectrum. In general, we find that the morphology of the gas is important to determine the power spectrum shape, in good agreement with \citet{Koch2020}. 

The lack of a correlation between the power spectrum slopes of the WISDOM galaxies and their SFRs matches the results in the literature. \citet{Zhang2012} found a tentative anti-correlation between SFR and power spectrum slope for dwarf galaxies with an absolute $B$-band magnitude lower than -14.5, but none for the entire sample of dwarf galaxies. Similarly, \citet{Dutta2009a} found tentative evidence for a correlation of power spectrum slope with SFR for five dwarf galaxies, but only scatter for a larger sample of 18 late-type galaxies \citep{Dutta2013}. 

The trend of steeper power spectrum slopes with increasing mean central molecular gas mass surface densities for the WISDOM galaxies seems somewhat more at odds with past results. However, while \citet{Koch2020} argued that the shallow power spectrum they measured for M~33 is caused by the central concentration of CO (within the inner 2 kpc), the behaviour of the WISDOM galaxies matches that of the \citet{Walker2014} simulated galaxies. In the latter, both of the galaxies with an exponential \hi mass surface density profile have a power spectrum steeper than that of the galaxy that evolved to have a flat \hi mass surface density profile (from the same initial conditions; \citealt{Walker2014}). Furthermore, \citet{Davis2022} found that the mean central molecular gas mass surface density is one of the best predictors (next to $\mus$) for the morphology of the central ISM.

Lastly, as previously highlighted by \citet{Koch2020}, we find that the observational parameters are important. The power spectrum slopes have a trend with the sensitivities of our observations, which may also affect the correlation with the central molecular gas mass surface densities (through its impact on the measured molecular gas masses). To further quantify the impact of observational parameters on the measured power spectrum slope, we performed tests with the simulations and refer to Appendices~\ref{A:obseff} and~\ref{A:simtests} for more detail. Even though the synthesised beam sizes (zeroth-moment map spatial resolutions), inclinations and mass surface density thresholds do not qualitatively affect the trend with the central stellar mass surface densities of the simulated galaxies' power spectra, they do make small quantitative differences. Similarly, the discrepancy of the power spectrum slopes when smoothing the simulated maps with Gaussian beams or mock interferometric beams highlights how sensitive the power spectrum slope is to how the data were obtained and the data analysis methods. This makes it doubly difficult to compare to literature results, and draw sound conclusions, as some discrepancies might be purely methodology dependent \citep[see also][]{Zhang2012,Koertgen2021}.  

\section{Summary and Conclusions} \label{s:Conclusion}
In this paper, we have analysed the cold gas mass surface density power spectra of the circumnuclear gas discs of 14 isolated, simulated galaxies from \citet{Gensior2020} and 12 galaxies from the WISDOM project \citep[e.g.][]{Onishi2017_WISDOMI_NGC3665}, making it the largest sample of CO extragalactic power spectra studied to date. The morphologies of the simulated galaxies range from pure disc to spheroid, with a fixed initial gas-to-stellar mass ratio of 5~per~cent, and the circumnuclear gas discs become increasingly smooth with increasing central stellar mass surface density ($\mus$), i.e. as the bulge dominance increases. The WISDOM galaxies comprise a mix of late- and early-type galaxies, some of which have very sub-structured, and others very smooth gas reservoirs. We computed the azimuthally-averaged, 1D power spectra of the gas mass surface densities, which are sensitive to the turbulent forcing of the gas. Our main findings are summarised below.

\begin{enumerate}[leftmargin=0.5cm]
    \item There is a strong correlation between the power spectrum slopes ($\beta$) and the central stellar mass surface densities of the simulated galaxies; the power spectra being steeper in galaxies with higher $\mus$ (and smoother central gas reservoirs). This confirms that the shear induced by the gravitational potential leads to dynamical suppression of fragmentation. 
    
    \item The flattening of the power spectrum slopes at $\log(\mus/\Msun~\kpc^{-2}) \geq 8.75$ (to a constant $\sim$2.6) could be physical or could be an artefact caused by the limited spatial resolution of the simulations and other parameter choices. If the flattening is real, the close agreement with the slope of two-dimensional \citet{Kolmogorov1941} turbulence (8/3) could indicate that high shear eventually drives incompressible turbulence. 
    
    \item Contrary to the simulations, the observations do not show any trend between power spectrum slope and central stellar mass surface density, whether they are being considered as a whole, or are divided into sub-samples of early- and late-type galaxies. However, the majority of the WISDOM galaxies have $\log(\mus/\Msun~\kpc^{-2}) \geq 8.75$, and have power spectrum slopes that scatter about the roughly constant slope of the simulations at such high $\mus$. This suggests that the flattening to a constant $\beta$ could be real and thus physical in origin. The large scatter of the observational slopes likely results from a combination of physical effects not captured by the simulations, the heterogeneity of the observations and the different gas morphologies (e.g. rings and spiral arms) affecting the (quality of the) power spectrum fits. 
    
    \item Of all the observation and galaxy properties we investigated, we only found a statistically significant correlation between the power spectrum slopes and the mean central molecular gas mass surface densities of the WISDOM galaxies. Although this correlation may depend on the sensitivities of the observations, this suggests that the dominant driver of turbulence in the ISM is related to the gas density, while the gravitational potential (of the stars) might play a less important role. 
    
\end{enumerate}

We have thus demonstrated using isolated galaxy simulations that the shear and turbulence induced by a spheroidal component that dominates the gravitational potential can suppress fragmentation in an embedded cold gas disc, resulting in a steep mass surface density power spectrum. The observations of 12 WISDOM galaxies however reveal a more nuanced reality, and the galaxies' power spectrum slopes depend most strongly on the mean central molecular gas mass surface densities. However, the heterogeneity of the observations and the occasionally poor description of a galaxy's power spectrum by a single power-law make it difficult to reach any strong conclusion. Velocity power spectra of the WISDOM galaxies may further elucidate the turbulence mechanisms and explain the different power spectrum slopes of galaxies with similar properties. Power spectra of a larger, more homogeneous sample of observed galaxies are required to firmly confirm whether the tentative correlation we discovered for the WISDOM sample galaxies holds true. In addition, high resolution cosmological (zoom) simulations, that resolve the cold ISM in sufficient detail, could shed more light on mass surface density power spectra from a theoretical perspective. 

\section*{Acknowledgements}

We thank Volker Springel for allowing us access to {\sc{Arepo}} and an anonymous referee for their helpful and constructive report. JG would like to thank Jonathan Henshaw, Maya Petkova, Luigi Bassini and Mauro Bernardini for helpful discussions. 
The authors acknowledge support by the High Performance and Cloud Computing Group at the Zentrum f{\"u}r Datenverarbeitung of the University of T{\"u}bingen, the state of Baden-W{\"u}rttemberg through bwHPC and the German Research Foundation (DFG) through grant no INST 37/935-1 FUGG. JG gratefully acknowledges financial support from the Swiss National Science Foundation (grant no CRSII5\_193826). JG and JMDK gratefully acknowledge funding from the Deutsche Forschungsgemeinschaft (DFG, German Research Foundation) through an Emmy Noether Research Group (grant number KR4801/1-1), as well as funding from the European Research Council (ERC) under the European Union's Horizon 2020 research and innovation programme via the ERC Starting Grant MUSTANG (grant agreement number 714907). JMDK gratefully acknowledges funding from the DFG Sachbeihilfe (grant number KR4801/2-1). COOL Research DAO is a Decentralized Autonomous Organization supporting research in astrophysics aimed at uncovering our cosmic origins. TAD and IR acknowledge support from the UK Science and Technology Facilities Council through grant ST/S00033X/1 and ST/W000830/1. MB was supported by STFC consolidated grant  ‘Astrophysics  at  Oxford’  ST/H002456/1 and ST/K00106X/1.
This paper makes use of the following ALMA data: ADS/JAO.ALMA 2013.1.00493.S, 2015.1.00419.S, 2015.1.00466.S, 2015.1.00597.S, 2016.1.00419.S, 2016.1.00437.S, 2016.2.00053.S, 2017.1.00391.S, 2017.1.00572.S, 2017.1.00904.S and 2017.1.00907.S. ALMA is a partnership of ESO (representing its member states), NSF (USA) and NINS (Japan), together with NRC (Canada), MOST and ASIAA (Taiwan), and KASI (Republic of Korea), in cooperation with the Republic of Chile. The Joint ALMA Observatory is operated by ESO, AUI/NRAO, and NAOJ. This research has made use of the NASA/IPAC Extragalactic Database (NED), which is operated by the Jet Propulsion Laboratory, California Institute of Technology, under contract with the National Aeronautics and Space Administration.  This  paper has also made use of the HyperLeda database (\url{http://leda.univ-lyon1.fr})

\section*{Data availability}
The raw observational data underlying this article is available from the ALMA archive at \url{https://almascience.eso.org/aq} using the project codes listed above. The data supporting the plots in this article will be shared on reasonable request to the corresponding author.

%%%%%%%%%%%%%%%%%%%%%%%%%%%%%%%%%%%%%%%%%%%%%%%%%%



\bibliographystyle{mnras}
\bibliography{references}

%%%%%%%%%%%%%%%%%%%%%%%%%%%%%%%%%%%%%%%%%%%%%%%%%%

%%%%%%%%%%%%%%%%% APPENDICES %%%%%%%%%%%%%%%%%%%%%

\appendix

\section{Impact of observational parameters on the power spectrum} \label{A:obseff}
The aim of this appendix is to quantify potential biases of observational limitations, such as the non-flux conserving radio beam, inclination of the galaxy and resolution on the  1D power spectrum and especially the slope of the power-law fit, $\beta$. To do so, we use the simulated galaxies, where we know the exact underlying gas density distributions, their 1D power spectra with the default settings used throughout the paper and can easily change the parameters. 

\subsection{Effect of the interferometric beam}\label{As:KinMS}
\begin{figure*}
    \centering
    \includegraphics[width=1.\linewidth]{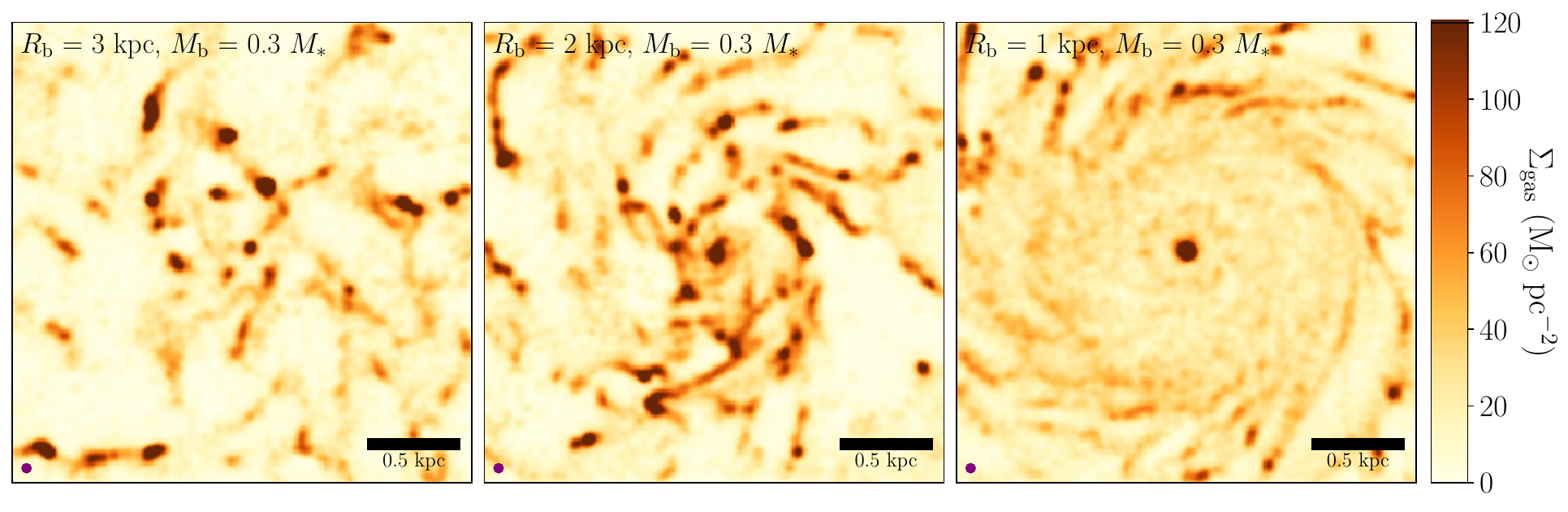}
    \caption{Total gas mass surface density maps generated using the \texttt{KinMS} tool, for the three galaxies with a bulge stellar mass of 0.3$\Mstar$ (see the top row of Figure~\ref{fig:gasSD_sims}), 600 Myr after the start of the simulations. Each map has a linear extent of 2.25 kpc and the solid purple circle in the bottom-left corner indicates the size of the synthesised beam. There is good agreement between the {\sc{Arepo}}-generated mass surface density projections and the \texttt{KinMS} maps, although the mock observations with \texttt{KinMS} tend to wash out small-scale structures.}
    \label{fig:KinMS_gasSD}
\end{figure*}

\begin{figure}
    \centering
    \includegraphics[width=1.\linewidth]{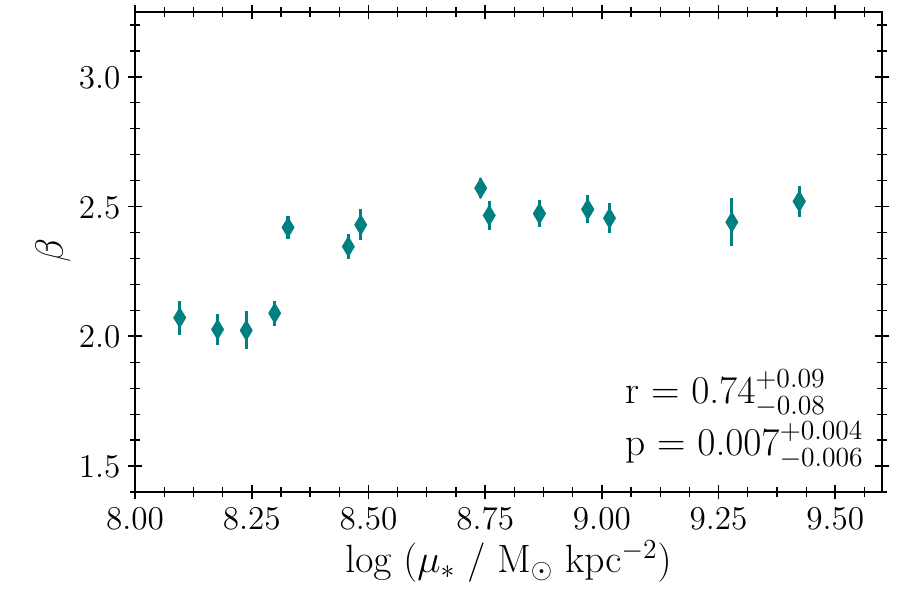}
    \caption{Time-averaged slopes ($\beta$) of the power-laws best-fitting the power spectra of the \texttt{KinMS} gas maps as a function of the galaxies' central stellar mass surface densities ($\mus$). Error bars indicate the uncertainties on the time-averaged power-law slopes. Although the dynamic range of the slopes is smaller than that in Figure~\ref{fig:pspec_slopes_mus_sims}, as indicated by the Spearman rank correlation coefficient and p-value listed in the bottom-right corner, the strong correlation between power spectrum slope and central stellar mass surface density persists.}
    \label{fig:KinMS_slopes_mustar}
\end{figure}

To test which effect a non-flux conserving radio beam has on the power spectrum, we use the \texttt{KinMS} tool to create mock interferometric observations from the gas particle distribution of all simulation snapshots used in the main analysis. \texttt{KinMS} can mimic observational effects, that are not taken into account when computing mass surface densities using {\sc{Arepo}}'s ray tracing method, such as beam smearing and velocity binning \citep{Davis2013a}. Although \texttt{KinMS} can model asymmetric synthesised beams at arbitrary position angles, as well as inclined gas distributions, we restrict ourselves here to the simple case of a symmetric beam with a FWHM of 43.3 pc and a disc inclination of 0$^{\circ}$, to match the simulation results presented in Section~\ref{ss:PS_sims}. We will explore the impact of beam size and inclination on the power spectra in Appendices~\ref{As:BS} and ~\ref{As:Inc}, respectively.

We pass \texttt{KinMS} the position and mass of the gas cells, as well as the circular rotation curve, of each simulated galaxy. The data cubes have a channel width of 10 kms$^{-1}$, a bandwidth of $\approx$1000 kms$^{-1}$ and a pixel size of 11 pc (corresponding to linearly sampling the synthesised beam with $\approx$4 pixels). As for the {\sc{Arepo}} mass surface density projections, we generate mock observations for 8 snapshots from 300 Myr to 1 Gyr. For illustration, Figure~\ref{fig:KinMS_gasSD} shows the maps of the three galaxies whose bulges contain 30~per~cent of the initial stellar mass at 600 Myr. Overall there is good agreement between the {\sc{Arepo}} (left column of Figure~\ref{fig:gasSD_sims}) and \texttt{KinMS} (Figure~\ref{fig:KinMS_gasSD}) maps. 

Figure~\ref{fig:KinMS_slopes_mustar} shows the time-averaged slops ($\beta$) of the power-laws best-fitting the power spectra, obtained from \texttt{KinMS} moment zero maps, of the simulated galaxies as a function of the central stellar mass surface densities of the corresponding simulations. Qualitatively we find the same trend of steepening power spectrum slope with increasing central stellar mass surface density for $\log(\mus/\Msun \kpc^{-2}) \leq 8.75$ that flattens for $\log(\mus/\Msun \kpc^{-2}) > 8.75$, as for the straight surface density projections (Figure~\ref{fig:pspec_slopes_mus_sims}). The Spearman rank correlation coefficient $r = 0.74^{+0.09}_{-0.08}$ with $p = 0.007 ^{+0.004}_{-0.006}$ indicates a strong and statistically-significant correlation. 
Quantitatively there are small differences between the power spectrum slopes of the galaxies with $\log(\mus/\Msun \kpc^{-2}) \leq 8.35$. Although these simulated galaxies show the same trend qualitatively, their \texttt{KinMS} power spectra have consistently slightly steeper slopes (2.01 -- 2.09) than those of the straight mass surface density projections (1.77 -- 1.89). This is the result of the synthesised \texttt{KinMS} non-normalised interferometric beam smoothing out some of the small-scale structures, thereby steepening the power spectrum slopes. However, we can reproduce the steeper slopes of the low-$\mus$ simulated galaxies (within the uncertainties) using the original, non-convolved, {\sc{Arepo}} ray-tracing maps and convolving them with the \texttt{KinMS} beam kernel. Since both methods yield qualitatively the same results, in this paper we use the {\sc{Arepo}} mass surface density projections convolved with a Gaussian kernel, which stay truer to the underlying gas distributions. However, this implies that the power spectra of interferometric observations could be artificially steepened by suppressing power on small scales. 

\subsection{Gaussian FWHM}\label{As:BS}
The beam of observations and the smoothing kernel used to generate the maps of the simulated galaxies can affect the shape of a power spectrum up to a spatial scale equal to a few times the FWHM$_{\rm Gauss}$ \citep[e.g.][]{Grisdale2017,Koch2020}. When fitting the power spectra, we therefore only consider spatial scales in excess of 3FWHM$_{\rm Gauss}$/$\sqrt{8\log2}$, which should mitigate this effect. Nonetheless, we test the robustness of our results to changes in Gaussian smoothing kernel size here, using the simulated galaxies. The effects should be similar when varying the synthesised beam size of a real observation. We vary the FWHM$_{\rm Gauss}$ we apply to the gas mass surface density projections from 20 to 70 pc, in 10 pc increments and subsequently re-grid each map such that FWHM$_{\rm Gauss}$ is always sampled with approximately 4 pixels. When fitting the power spectra, we account for the differences of the Gaussian FWHM by adjusting the lower spatial limit of the fits accordingly. 

\begin{figure}
    \centering
    \includegraphics[width=1.\linewidth]{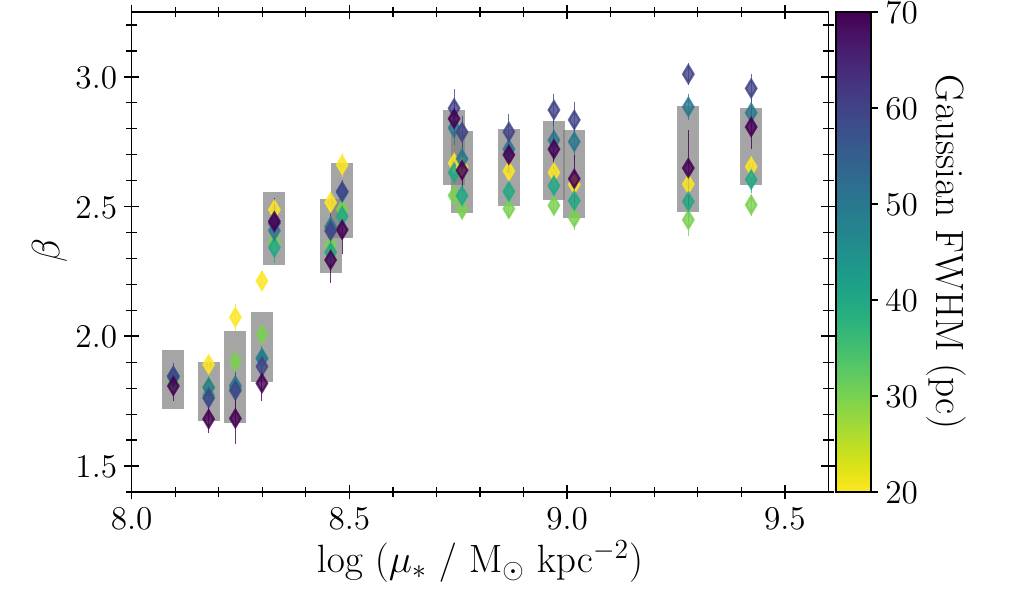}
    \caption{Time-averaged slopes ($\beta$) of the power-laws best-fitting the power spectra of the simulated galaxies as a function of the galaxies' central stellar mass surface densities ($\mus$). The data points are colour-coded by the Gaussian FWHM of the smoothing kernels with which the maps were spatially smoothed prior to computing the power spectra. The error bars indicate the uncertainties on the time-averaged power-law slopes while the grey-shaded regions indicate the average uncertainties of the power-law fits.}
    \label{fig:beamsize_test}
\end{figure}

Figure~\ref{fig:beamsize_test} shows the time-averaged power spectrum slopes of the simulated galaxies as a function of their central stellar mass surface densities, colour-coded by the Gaussian FWHM sizes. The error bars denote the uncertainty on the time-averaged best-fitting power-law slopes, while the grey-shaded regions indicate the average uncertainty of the individual power-law fits. There are some changes of the power spectrum slope with varying FWHM$_{\rm Gauss}$ size, however these are mostly over slope ranges comparable to the uncertainty of the power-law fits and there is no clear pattern of how the FWHM$_{\rm Gauss}$ size affects the power spectrum slope.

\begin{table}
    \centering
    \caption{Spearman rank correlation coefficients and p-values of the correlations between power spectrum slope and central stellar mass surface density of the simulated galaxies, when using different Gaussian FWHM for the spatial smoothing kernel applied to the gas mass surface density maps.}
    \renewcommand{\arraystretch}{1.5}
    \begin{tabular}{ccc}
        \hline
        Gaussian FWHM & Spearman r & $\log$ p \\
        (pc) & & \\
        \hline
        20 & 0.74$_{-0.07}^{+0.07}$ & -2.34$_{-0.97}^{+0.28}$ \\
        30 & 0.76$_{-0.07}^{+0.07}$ & -2.47$_{-1.13}^{+0.28}$ \\
        40 & 0.82$_{-0.05}^{+0.05}$ & -3.17$_{-1.16}^{+0.22}$ \\
        50 & 0.92$_{-0.04}^{+0.03}$ & -4.74$_{-1.81}^{+0.24}$ \\
        60 & 0.92$_{-0.03}^{+0.03}$ & -4.89$_{-1.89}^{+0.19}$ \\
        70 & 0.82$_{-0.05}^{+0.05}$ & -3.12$_{-1.12}^{+0.22}$ \\
        \hline
    \end{tabular}
    \label{tab:beamsize_rp}
\end{table}

We list the Spearman rank correlation coefficient and the probability that the trend of each Gaussian kernel data set is statistically significant in Table~\ref{tab:beamsize_rp}. For all FWHM$_{\rm Gauss}$ considered here, the correlation between power spectrum slope and central stellar mass surface density remains strong and statistically significant. In short, although there are some small quantitative changes of power spectrum slope as a function of the Gaussian smoothing kernel size, these differences are only of the order of the uncertainties of the fits. Qualitatively, at a given $\mus$, there is good agreement between all the power spectra.

\subsection{Inclination}\label{As:Inc}
All selected WISDOM galaxies have inclinations ($i$) ranging from 20 to 70$^{\circ}$. The power spectra are computed from (molecular) gas mass surface density maps, and a large part of the analysis focusses on the central stellar mass surface densities of the galaxies, i.e. using projected quantities. Therefore, the inclination of the galaxies could affect both the 1D power spectra (and thus $\beta$), as well as $\mus$. Here, we systematically test the impact of inclinations ranging from 0 to 90$^{\circ}$, in 10$^{\circ}$ increments, on our simulated galaxy results, by repeating the analysis after inclining the simulated galaxies. We use \texttt{KinMS} to produce the inclined total gas mass surface density projections adopting the same set-up described in Appendix~\ref{As:KinMS}. We also re-compute the stellar half-mass radius and $\mus$ for each inclined stellar particle distribution. 

\begin{figure}
    \centering
    \includegraphics[width=1.\linewidth]{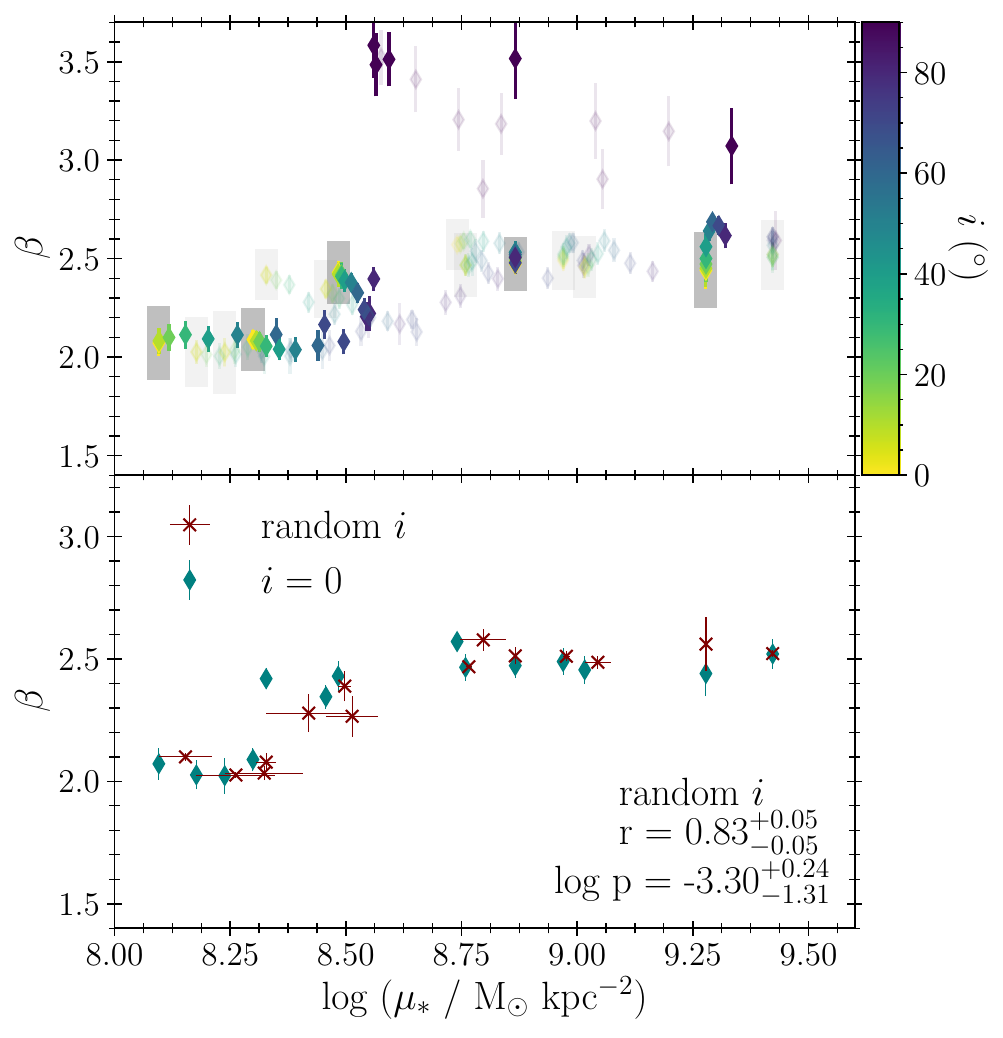}
    \caption{Time-averaged slopes ($\beta$) of the power-laws best-fitting the power spectra of the simulated galaxies as a function of the galaxies' central stellar mass surface densities ($\mus$). Top panel: the data points are colour-coded by the inclinations ($i$) of the simulated galaxies, only showing every third data point with full opacity.% for better visibility of trends. 
    The error bars indicate the the uncertainties on the time-averaged power-law slopes while the grey-shaded regions indicate the average uncertainties of the power-law fits. Bottom panel: The red crosses show the median $\beta$ and $\mus$ for $i$ ranging from 0 - 70$^{\circ}$, error bars indicating the 16th-to-84th percentile variation determined via bootstrapping, overplotted on the $i=0$ data points (teal diamonds).}
    \label{fig:inc_test}
\end{figure}

The top panel of Figure~\ref{fig:inc_test} shows the time-averaged power spectrum slopes of the simulated galaxies as a function of their central stellar mass surface densities, colour-coded by the inclination of the simulated galaxy disc from which the gas mass surface density map was produced. The error bars denote the uncertainty on the time-averaged best-fitting power-law slopes, while the grey-shaded regions indicate the average uncertainty of the individual power-law fits. We only show the points for every third galaxy at full opacity for better readability of the plot. For all galaxies with B/T < 1  $\mus$ increases with $i$, by up to 0.5 dex for the disc-dominated cases and by $\lesssim$0.15 dex for the more massive and compact bulges. The best-fitting power-law slope also changes with inclination. However, apart from the extreme, $i=90^{\circ}$ case, the slopes are generally consistent with each other within the uncertainties on the time-averaged slopes, and they are always within the uncertainties of the individual power-law fits. The correlation between $\beta$ and $\mus$ is preserved for $i\leq80^{\circ}$, it even becomes stronger for $i\leq60$ ($r = 0.90^{+0.04}_{-0.04}$ with $\log p = -4.242_{-1.746}^{+0.156}$ cf. $r = 0.74^{+0.09}_{-0.08}$ with $p = 0.007 ^{+0.004}_{-0.006}$ for $i=0^{\circ}$), due to the increase of $\mus$ with $i$. By contrast, when the simulated galaxies are viewed edge-on, all power spectra are very steep. For low-$\mus$ galaxies, some steepening already occurs at $i\geq 70^{\circ}$. Thus, caution is advised when considering the power spectra of galaxies whose gas reservoirs have inclinations above 70$^{\circ}$. This is in good agreement with \citet{Grisdale2017}, who compared the power spectra of their simulated galaxies at $i=0^{\circ}$ and $i=40^{\circ}$, and found that the main difference is the total power in the power spectra, while the power-law slopes change by very little. 

Since the WISDOM galaxies to which we compare the simulated galaxies span a range of inclinations, we perform a second test where we use bootstrapping to determine median $\beta$ and $\mus$, as well as their 16th-to-84th percentile variations, randomly drawing an inclination between $0-70^{\circ}$ each time. This is shown as the dark red crosses in the bottom panel of Figure~\ref{fig:inc_test}. The resultant trend of increasing $\beta$ with $\mus$ (at $\log (\mus/\Msun~\kpc^{-2}) \leq 8.75$) is more pronounced for the bootstrapped data, and has a Spearman rank correlation coefficient $r = 0.83^{+0.05}_{-0.05}$ with a p-value $p = 0.0005 ^{+0.009}_{-0.00002}$. This further confirms that the power-law slope and the trend the simulations show with $\mus$ is not strongly affected for $i\lesssim70^{\circ}$. 

\section{Simulation parameter tests}\label{A:simtests}
 
\subsection{Surface density threshold}\label{As:SD_thresh}
We use the total gas mass surface density projections of the simulated galaxies' gas discs to compute the power spectrum in the main analysis, as we do not use a chemical model that models molecular gas or tracks CO specifically. In contrast, we use molecular gas mass surface density maps to compute power spectra for the observed galaxies. Here, we therefore quantify how the best-fitting power-law slopes of the power spectra are affected by using total gas mass surface density maps clipped to surface density thresholds ($\Sigma_{\rm thresh}$) of up to 100 $\Msun~\pc^{-2}$, to test how robust the results of the main analysis remain when using a density threshold that mimics molecular gas. We use the {\sc{Arepo}}-generated total gas mass surface density maps, masking all surface densities smaller than our threshold prior to computing the power spectra. 

\begin{figure}
    \centering
    \includegraphics[trim={0cm 0.3cm 0cm 0cm}, clip=true,width=1.\linewidth]{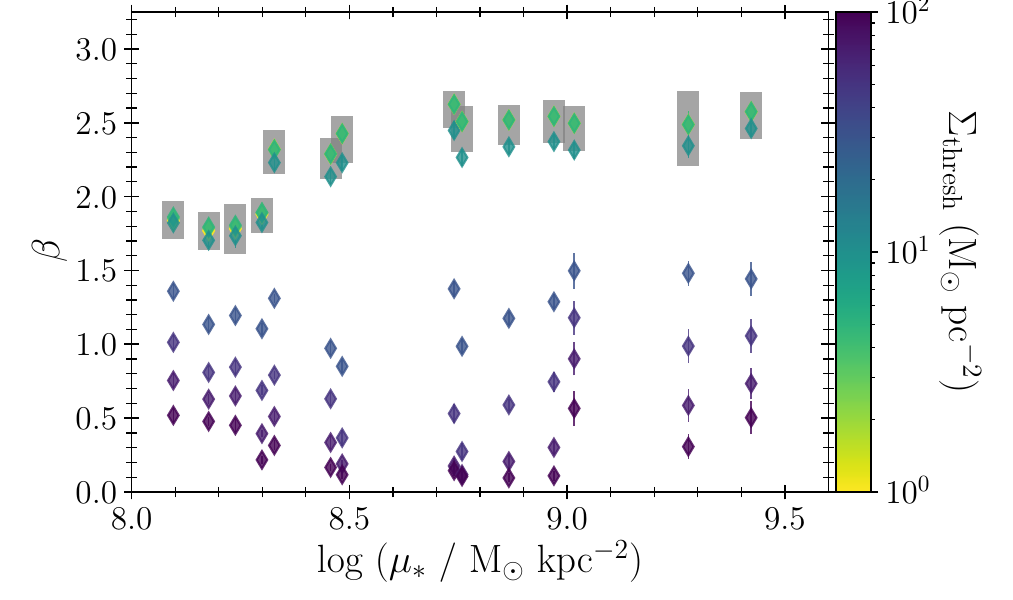}
    \caption{Time-averaged slopes ($\beta$) of the power-laws best-fitting the power spectra of the simulated galaxies as a function of the galaxies' central stellar mass surface densities ($\mus$). The data points are colour-coded by the gas mass surface density thresholds imposed on the maps prior to computing the power spectra. The error bars indicate the uncertainties on the time-averaged power-law slopes while the grey-shaded regions indicate the average uncertainties of the power-law fits. The data points with $\Sigma_{\rm thresh} = 0, 1, 3, 5~\Msun~\pc^{-2}$ systematically fall on top of each other.}
    \label{fig:SD_thresh_test}
\end{figure}

Figure~\ref{fig:SD_thresh_test} shows the time-averaged power spectrum slopes of the simulated galaxies as a function of their central stellar mass surface densities, colour-coded by the $\Sigma_{\rm thresh}$ of the gas mass surface density maps from which the power spectra were computed. The error bars denote the uncertainty on the time-averaged best-fitting power-law slopes, while the grey-shaded regions indicate the average uncertainty of the individual power-law fits. For $\Sigma_{\rm thresh}<10~\Msun~\pc^{-2}$, the power-law slopes remain unaffected, the data points systematically fall on top of each other. For $\Sigma_{\rm thresh}=10~\Msun~\pc^{-2}$, the power-law slopes are in good agreement with the ones of the unclipped maps, in some cases within the uncertainties of the time-averaged slopes and always within the average uncertainties of the individual fits. Notable differences are however present for $\Sigma_{\rm thresh} \geq 30~\Msun~\pc^{-2}$, i.e. for the darker blue and purple diamonds in Figure~\ref{fig:SD_thresh_test}. At those high $\Sigma_{\rm thresh}$, the power spectra become much shallower as the threshold is increased for all simulated galaxies. This is the result of ignoring low-density regions, or conversely focussing on increasingly small (very) high-density peaks. The latter are preferentially found in the low-$\mus$ galaxies, which have clumpier ISM with starker density contrasts. Therefore, the differences of the power-law slopes are larger for the high-$\mus$ galaxies. Having said that, these high surface density thresholds are somewhat unphysical for our simulations, as the gas that remains looks like a noise map in each case. Furthermore, most observations of CO in galaxies now routinly reach surface densities smaller than 10 $\Msun~\pc^{-2}$. As before, we list the Spearman rank correlation coefficients and the p-values of the correlations for different thresholds in Table~\ref{tab:SDthresh_rp}. In summary, our results are robust even when introducing a reasonable gas mass surface density threshold that mimics (high-density) molecular gas, such as $\Sigma_{\rm thresh}\approx10~\Msun~\pc^{-2}$.

\begin{table}
    \centering
    \caption{Spearman rank correlation coefficients and p-values of the correlations between power spectrum slope and central stellar mass surface density of the simulated galaxies, when using maps with different gas mass surface density thresholds.}
    \renewcommand{\arraystretch}{1.5}
    \begin{tabular}{ccc}
        \hline
        $\Sigma_{\rm thresh}$ & Spearman r & $\log$ p \\
        ($\Msun~\pc^{-2}$) & & \\
        \hline
        0.0 & 0.82$_{-0.05}^{+0.05}$ & -3.08$_{-1.08}^{+0.17}$ \\
        1.0 & 0.81$_{-0.06}^{+0.05}$ & -3.00$_{-1.09}^{+0.26}$ \\
        3.0 & 0.81$_{-0.05}^{+0.05}$ & -3.06$_{-1.10}^{+0.24}$ \\
        5.0 & 0.81$_{-0.05}^{+0.05}$ & -3.04$_{-1.13}^{+0.22}$ \\
        10.0 & 0.84$_{-0.05}^{+0.05}$ & -3.29$_{-1.31}^{+0.14}$ \\
        30.0 & 0.39$_{-0.09}^{+0.09}$ & -0.73$_{-0.38}^{+0.20}$ \\
        50.0 & 0.11$_{-0.07}^{+0.07}$ & -0.15$_{-0.11}^{+0.10}$ \\
        70.0 & -0.08$_{-0.09}^{+0.09}$ & -0.13$_{-0.13}^{+0.09}$ \\
        100.0 & -0.20$_{-0.10}^{+0.11}$ & -0.29$_{-0.25}^{+0.17}$ \\
        \hline
    \end{tabular}
    \label{tab:SDthresh_rp}
\end{table}

\subsection{Centre masking}\label{As:Cmask}
\begin{figure}
    \centering
    \includegraphics[trim={0cm 0.3cm 0cm 0cm}, clip=true,width=1\linewidth]{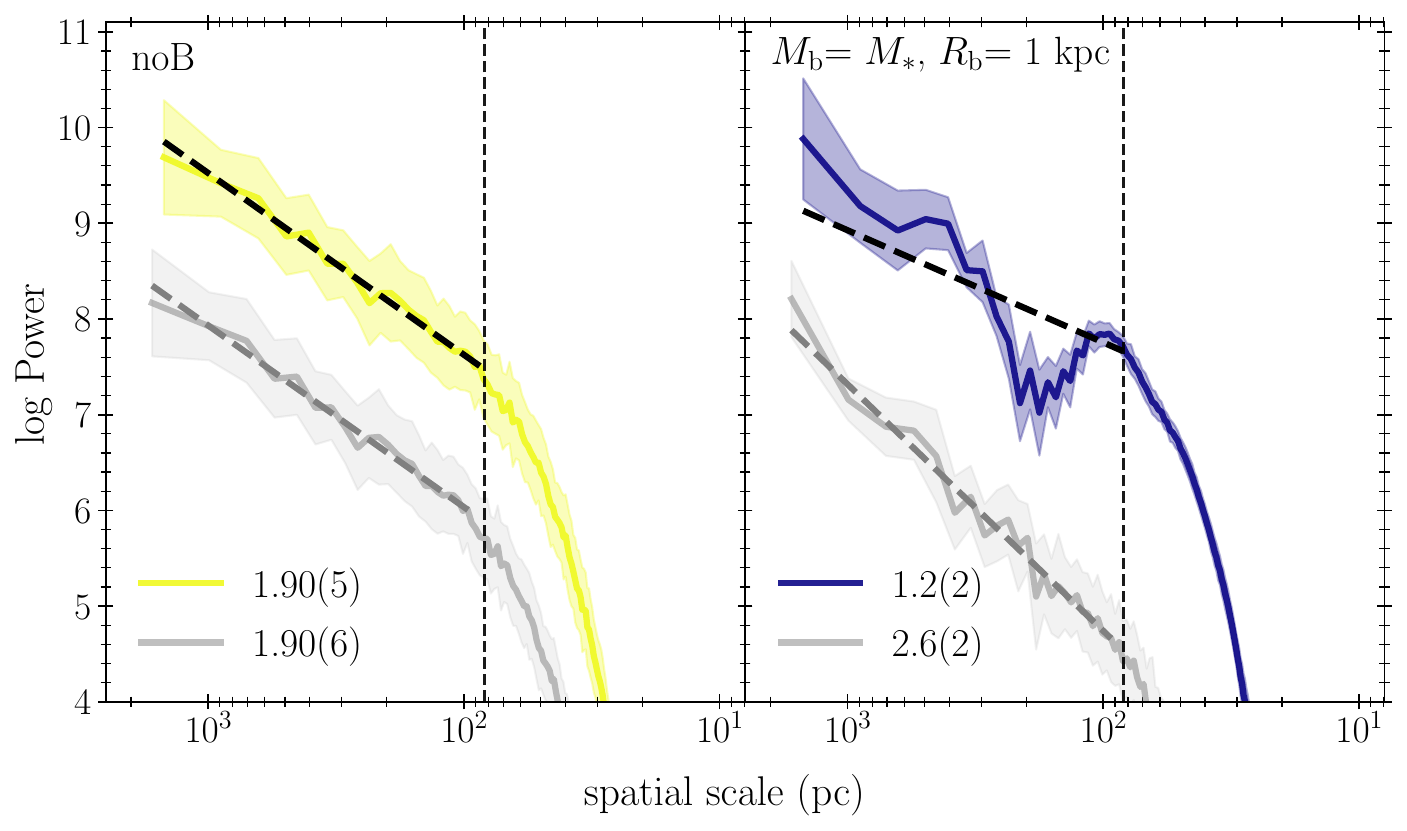}
    \caption{Gas mass surface density power spectra of the disc-only (noB, left panel) and most compact spheroid (B\_M100\_R1, right panel), without masking their centres (coloured lines, with uncertainties as shading) and, offset, with the central masking applied (grey lines, with uncertainties as shading). In each case, the best-fitting power-law is shown as a dashed line, the vertical black dashed line indicates the smallest spatial scale considered for the fit and the legend lists the best-fitting power-law slope with its 1$\sigma$ uncertainty. The non-masked power spectrum of B\_M100\_R1 clearly dominated by the very dense accumulation of gas at the centre of the galaxy (on scales up to 200 pc), while the power spectrum of noB is nearly identical to the centre-masked one.}
    \label{fig:ex_nomask_pspec}
\end{figure}

Lastly, we elaborate on the need for masking the central regions of the gas mass surface density maps of the simulated galaxies prior to our computation of the power spectrum. Figure~\ref{fig:ex_nomask_pspec} provides an example of just how extreme the effect of the very dense central regions in bulge-dominated galaxies can be, by contrasting the power spectra of the most disc-dominated (noB) and most bulge-dominated (B\_M100\_R1) simulated galaxies without central masking (colour) and with central masking (grey, the same power spectra as in Figure~\ref{fig:pspec_sims_3panel}). In the unmasked power spectrum of the bulge-dominated galaxy, there is a bump on the spatial scales associated with the central region (200 pc and smaller), preceded by an order of magnitude drop in power, that make describing the power spectrum with a power-law pointless. This central accumulation is a consequence of the suppression of star formation in the galactic centre and the lack of feedback from an active galactic nucleus and is therefore unphysical (see also discussion in \citealt{Gensior2020}). To avoid contamination of the power spectra by such bright centres, we simply mask these regions in the gas mass surface density maps using an inverted cosine bell window function of \texttt{turbustat}. The masking function naturally tapers off, so that there is a smooth transition between the completely masked centre and the rest of the surface density maps, thus avoiding Gibbs ringing artefacts that would bias the power spectra. The masking function is characterised by the masking parameter $\alpha$, where a smaller $\alpha$ corresponds to a smaller region being masked. We choose $\alpha=0.3$ for the simulated galaxies, which fully masks the gas accumulations in the central regions of the bulge-dominated galaxies and results in power spectra that are well-described by a single power-law (grey cf. dark blue power spectra in the right panel of Figure~\ref{fig:ex_nomask_pspec}). By contrast, the centre masking has very little effect on the power spectra and their best-fitting power-law slopes of the disc-dominated galaxies that have a much clumpier ISM (grey cf. yellow power spectra in the left panel of Figure~\ref{fig:ex_nomask_pspec}). For the observations, we adjust $\alpha$ for each of the selected WISDOM galaxies, such that the masked region snugly fits the bright central CO emission.

\section{WISDOM power spectrum slope (non-)correlations} \label{A:Corr}
\begin{table*}
    \renewcommand{\arraystretch}{1.5}
    \centering
    \caption{Spearman rank correlation coefficients and p-values of the correlations between the power spectrum slope and assorted properties of the selected WISDOM galaxies.}
    \makebox[\textwidth][c]{
\begin{tabular}{lcccccc}
    \hline
    & \multicolumn{2}{c}{All galaxies} & \multicolumn{2}{c}{ETGs} & \multicolumn{2}{c}{LTGs} \\
    \hline
    Quantity & Spearman r & p-value & Spearman r & p-value & Spearman r & p-value \\
    \hline
    $\bar{\Sigma}_{\rm H_2}$ & 0.61$^{+0.13}_{-0.13}$ & 0.06$^{+0.05}_{-0.05}$ & 0.63$^{+0.15}_{-0.16}$ & 0.13$^{+0.11}_{-0.11}$ & 0.45$^{+0.35}_{-0.45}$ & 0.41$^{+0.46}_{-0.31}$ \\
    $M_{\rm H_2}$ & 0.28$^{+0.17}_{-0.17}$ & 0.41$^{+0.30}_{-0.27}$ & 0.30$^{+0.27}_{-0.26}$ & 0.49$^{+0.27}_{-0.31}$ & 0.12$^{+0.48}_{-0.52}$ & 0.59$^{+0.21}_{-0.39}$\\
    SFR & 0.35$^{+0.17}_{-0.16}$ & 0.31$^{+0.25}_{-0.23}$ & 0.30$^{+0.22}_{-0.26}$ & 0.46$^{+0.31}_{-0.28}$ & 0.19$^{+0.41}_{-0.39}$ & 0.59$^{+0.28}_{-0.31}$ \\
    sSFR & 0.44$^{+0.19}_{-0.18}$ & 0.22$^{+0.20}_{-0.19}$ & 0.33$^{+0.24}_{-0.24}$ & 0.44$^{+0.34}_{-0.30}$ & 0.37$^{+0.33}_{-0.37}$ & 0.49$^{+0.38}_{-0.30}$ \\
    SFE & 0.12$^{+0.21}_{-0.21}$ & 0.59$^{+0.28}_{-0.30}$ & 0.20$^{+0.30}_{-0.31}$ & 0.56$^{+0.32}_{-0.35}$ & -0.35$^{+0.55}_{-0.45}$ & 0.50$^{+0.30}_{-0.30}$ \\
    $f_{\rm H_2}$ & 0.40$^{+0.17}_{-0.16}$ & 0.25$^{+0.21}_{-0.20}$ & 0.41$^{+0.19}_{-0.20}$ & 0.39$^{+0.26}_{-0.24}$ & 0.24$^{+0.56}_{-0.64}$ & 0.55$^{+0.25}_{-0.35}$ \\
    $\bar{\Sigma}_{\rm H_2}$/$\mu_{\ast}$ & 0.41$^{+0.17}_{-0.16}$ & 0.24$^{+0.19}_{-0.19}$ & 0.48$^{+0.19}_{-0.19}$ & 0.27$^{+0.22}_{-0.20}$ & 0.20$^{+0.50}_{-0.40}$ & 0.55$^{+0.32}_{-0.36}$ \\
    $\mu_{\ast}$ & -0.14$^{+0.24}_{-0.24}$ & 0.55$^{+0.31}_{-0.34}$ & -0.09$^{+0.33}_{-0.31}$ & 0.57$^{+0.29}_{-0.31}$ & 0.02$^{+0.48}_{-0.52}$ & 0.57$^{+0.30}_{-0.39}$ \\
    $M_{\ast}$ & -0.22$^{+0.20}_{-0.19}$ & 0.49$^{+0.34}_{-0.32}$ & -0.16$^{+0.25}_{-0.25}$ & 0.62$^{+0.25}_{-0.30}$ & -0.51$^{+0.21}_{-0.29}$ & 0.39$^{+0.24}_{-0.28}$ \\
    $\Sigma_{\rm gas, thresh}$ & 0.52$^{+0.16}_{-0.17}$ & 0.14$^{+0.14}_{-0.13}$ & 0.58$^{+0.20}_{-0.20}$ & 0.18$^{+0.17}_{-0.16}$ & 0.40$^{+0.30}_{-0.30}$ & 0.49$^{+0.38}_{-0.30}$ \\
    $l_{\rm min}$ & 0.06$^{+0.15}_{-0.15}$ & 0.70$^{+0.21}_{-0.22}$ & 0.21$^{+0.17}_{-0.21}$ & 0.58$^{+0.24}_{-0.23}$ & 0.12$^{+0.28}_{-0.32}$ & 0.70$^{+0.17}_{-0.19}$ \\
    $l_{\rm max}-l_{\rm min}$ & 0.15$^{+0.17}_{-0.17}$ & 0.60$^{+0.28}_{-0.28}$ & 0.10$^{+0.19}_{-0.19}$ & 0.68$^{+0.23}_{-0.23}$ & 0.47$^{+0.33}_{-0.47}$ & 0.42$^{+0.45}_{-0.31}$ \\
    Beam size & 0.06$^{+0.15}_{-0.15}$ & 0.70$^{+0.22}_{-0.24}$ & 0.21$^{+0.17}_{-0.21}$ & 0.58$^{+0.24}_{-0.23}$ & 0.12$^{+0.28}_{-0.32}$ & 0.69$^{+0.18}_{-0.19}$ \\
    Gini & 0.04$^{+0.27}_{-0.26}$ & 0.57$^{+0.30}_{-0.30}$ & 0.05$^{+0.35}_{-0.37}$ & 0.57$^{+0.30}_{-0.32}$ & -0.49$^{+0.39}_{-0.41}$ & 0.40$^{+0.34}_{-0.36}$ \\
    Smoothness & 0.03$^{+0.19}_{-0.20}$ & 0.67$^{+0.25}_{-0.24}$ & -0.10$^{+0.24}_{-0.22}$ & 0.70$^{+0.24}_{-0.22}$ & -0.69$^{+0.29}_{-0.21}$ & 0.23$^{+0.27}_{-0.20}$ \\
    Asymmetry & 0.03$^{+0.23}_{-0.23}$ & 0.62$^{+0.26}_{-0.26}$ & -0.19$^{+0.15}_{-0.17}$ & 0.65$^{+0.23}_{-0.22}$ & -0.47$^{+0.27}_{-0.33}$ & 0.42$^{+0.33}_{-0.31}$ \\
    $Q_{\ast}/Q_{\rm gas}$ & 0.28$^{+0.24}_{-0.24}$ & 0.43$^{+0.34}_{-0.32}$ & 0.13$^{+0.47}_{-0.50}$ & 0.50$^{+0.38}_{-0.39}$ & 0.13$^{+0.67}_{-0.53}$ & 0.57$^{+0.23}_{-0.37}$\\
    \hline
 \multicolumn{7}{p{0.63\linewidth}}{\footnotesize{\textit{Notes:} Mean Spearman rank correlation coefficients, p-values and their 16th and 84th percentiles have been estimated from 1000 Monte-Carlo simulations of the data, to take into account uncertainty on the power-law fit and the observational data. The correlation coefficients and p-values of the entire sample and sub-sets of early- and late-type galaxies are listed. We test for correlation with: the mean molecular gas mass surface density within the central kpc ($\bar{\Sigma}_{\rm H_2}$), the total molecular gas mass ($M_{\rm H_2}$), the star formation rate (SFR), the specific star formation rate (sSFR), the star formation efficiency (SFE=SFR/$M_{\rm H_2}$), the molecular gas-to-stellar total mass ratio ($f_{\rm H_2}$), the molecular gas-to-stellar central mass surface density ratio ($\bar{\Sigma}_{\rm H_2}$/$\mu_{\ast}$), the central stellar mass surface density ($\mu_{\ast}$), the total stellar mass ($M_{\ast}$), the molecular gas mass surface density sensitivity ($\Sigma_{\rm gas, thresh}$), the smallest spatial scale of the power-law fit ($l_{\rm min}$), the spatial scale extent of the power-law fit ($l_{\rm max}-l_{\rm min}$), the geometric average of the synthesised beam (Beam size), the non-parametric morphological indicators Gini, Smoothness and Asymmetry, and the stellar-to-gas \citet{Toomre1964} Q ratio.}}
    \end{tabular}
}
    \label{tab:obs_cor_all}
\end{table*}

We summarise our search for correlations between the power spectrum slopes of the WISDOM galaxies and various properties in Table~\ref{tab:obs_cor_all}. For each quantity we show the mean Spearman rank correlation coefficient and p-value and indicate their 16th and 84th percentiles from 1000 Monte-Carlo simulations, performed to account for the uncertainty of each power spectrum fit and the observational quantities. In addition to listing those two parameters for the entire sample, we also list them for the sub-samples of ETGs and LTGs. 

\section{Multi-component power-laws}\label{A:brokenpowerlaw}
As mentioned in Section~\ref{ss:PS_obs}, the power spectra of NGC 4501, NGC 4826 and NGC 5806 can also be fit with a dual-component (broken) power-law. Because the single power-law fit to the power spectra of these galaxies remains a reasonable fit, consistent with the power spectra to within the 1 $\sigma$ uncertainties, we have used the single power-law slopes for all observed galaxies in the main body of the paper. In this Appendix, we discuss the broken power-law fits to the power spectra of NGC 4501, NGC 4826 and NGC 5806 in more detail and include their large scale slopes in the key figures of this paper. 

\begin{figure*}
    \centering
    \includegraphics
    [width=.97\linewidth]{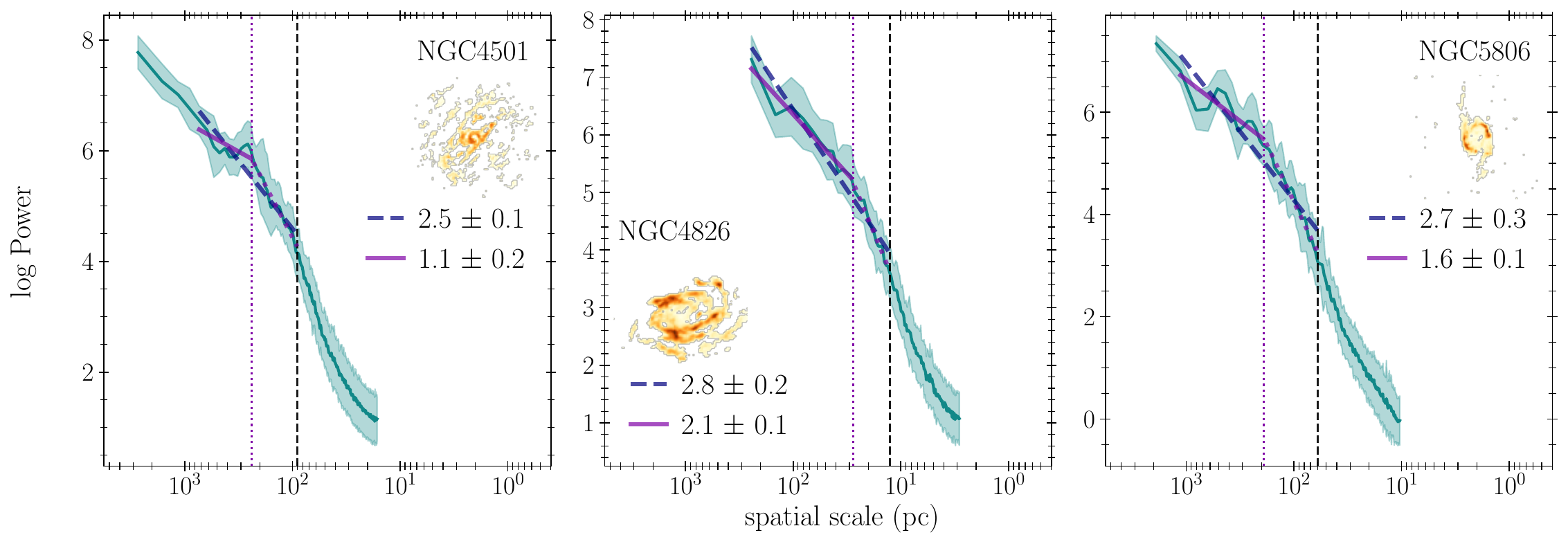}
    \caption{Gas mass surface density power spectra of the three WISDOM galaxies NGC 4501, NGC 4826 and NGC 5806 (solid teal lines, with uncertainties as shading), plotted with unique spatial scale for ease of comparison. In each panel, the best-fitting single power-law is shown as navy dashed line, the large scale (small scale) best-fitting broken power-law is shown as a purple solid (dotted) line, the vertical black dashed (purple dotted) line indicates the smallest spatial scale (break scale) considered for the fit and the legend lists the best-fitting (large scale) power-law slopes with their 1$\sigma$ uncertainty. The inset shows the centre-masked moment zero map of the galaxy, from which the power spectrum was computed.}
    \label{fig:pspec_obs_indv_brk}
\end{figure*}

Like for the single power-law fits, we use \texttt{turbustat} to fit a dual-component power-law to the 1D spatial power spectrum. The broken power-law model implemented in \texttt{turbustat} is a linear segmented model, which means the sum of two power-laws with a rigid break point. Specifically:
\begin{equation}
    P(k) = \begin{cases} Ak^{-\beta}  & k < k_{\rm brk} \\
                        Ak^{-\beta}\left(\frac{k}{k_{\rm brk}}\right)^{\beta - \beta_2} & k > k_{\rm brk}
    \end{cases}
\end{equation}\label{eq:linsegmodel}where $\beta$ is the slope of the large scale power-law, $k_{\rm brk}$ the break point, and $\beta_2$ the slope of the small scale power-law. Table~\ref{tab:obs_pspecfit_brk} lists the results of the best-fitting broken power-laws, which are overplotted on the 1D gas mass surface density power spectrum and the best-fitting single power-law in Figure~\ref{fig:pspec_obs_indv_brk}.\footnote{Fitting the 1D power spectra with a broken power-law like those presented in \citet{Koch2020} or \citet{Koertgen2021}, which allow for a smoother transition around the break scale, using \texttt{scipy.optimize.curve\_fit}, has a negligible impact on the the large-scale slope and break point of the best fitting dual-component power-law.}  

\begin{table}
\centering
\caption{Results of the broken power-law fits to the 1D power spectra of the WISDOM galaxies NGC 4501, NGC 4826 and NGC 5806.}
 \begin{tabular}{ccccc}
  \hline
  Galaxy & $\beta$ & $\beta_2$ & brk & Fit range\\
   & & & (pc) & (pc) \\
  \hline
NGC 4501 & $1.10 \pm 0.19$ & $3.74 \pm 0.34$ & 239 &  90 -- 816 \\
NGC 4826 & $2.05 \pm 0.12$ & $4.53 \pm 0.33$ & 27 & 13 -- 246 \\
NGC 5806 & $1.62 \pm 0.14$ & $4.47 \pm 0.30$ & 191 & 60 -- 1141 \\
  \hline
  \multicolumn{5}{p{0.85\linewidth}}{\footnotesize{\textit{Notes:} For each galaxy, column 1 lists the name, columns 2 and 3 list the respective best-fitting large and small scale slope of the dual power-law and the $1\sigma$ uncertainty. Column 4 lists the break point and column 5 lists the spatial scales across which a the power-law was fitted to the 1D power spectrum.}}
 \end{tabular}
 \label{tab:obs_pspecfit_brk}
\end{table}

\begin{figure}
    \centering
    \includegraphics[width=\linewidth]{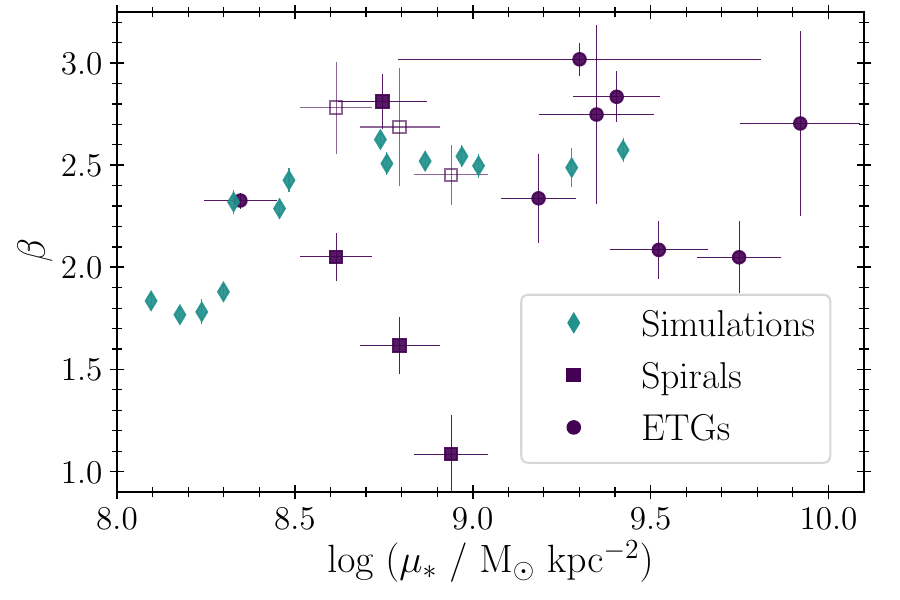}
    \caption{Slopes ($\beta$) of the power-laws best-fitting the power spectra of the simulated (teal) and WISDOM (purple) galaxies as a function of their central stellar mass surface densities ($\mus$), including the large scale slopes of the broken power-laws for NGC 4501, NGC 4826 and NGC 5806. The empty symbols show the result of the best-fitting single power-law to the aforementioned galaxies. The vertical error bars indicate the uncertainties on the time-averaged best-fitting power-law slopes (simulations) or the 1$\sigma$ uncertainties on the best-fitting power-law slopes (observations).}
    \label{fig:mus_slopes_wls}
\end{figure}

\begin{figure*}
    \centering
    \includegraphics[trim={2.2cm 1cm 0.5cm 0.5cm}, clip=true ,width=0.95\linewidth]{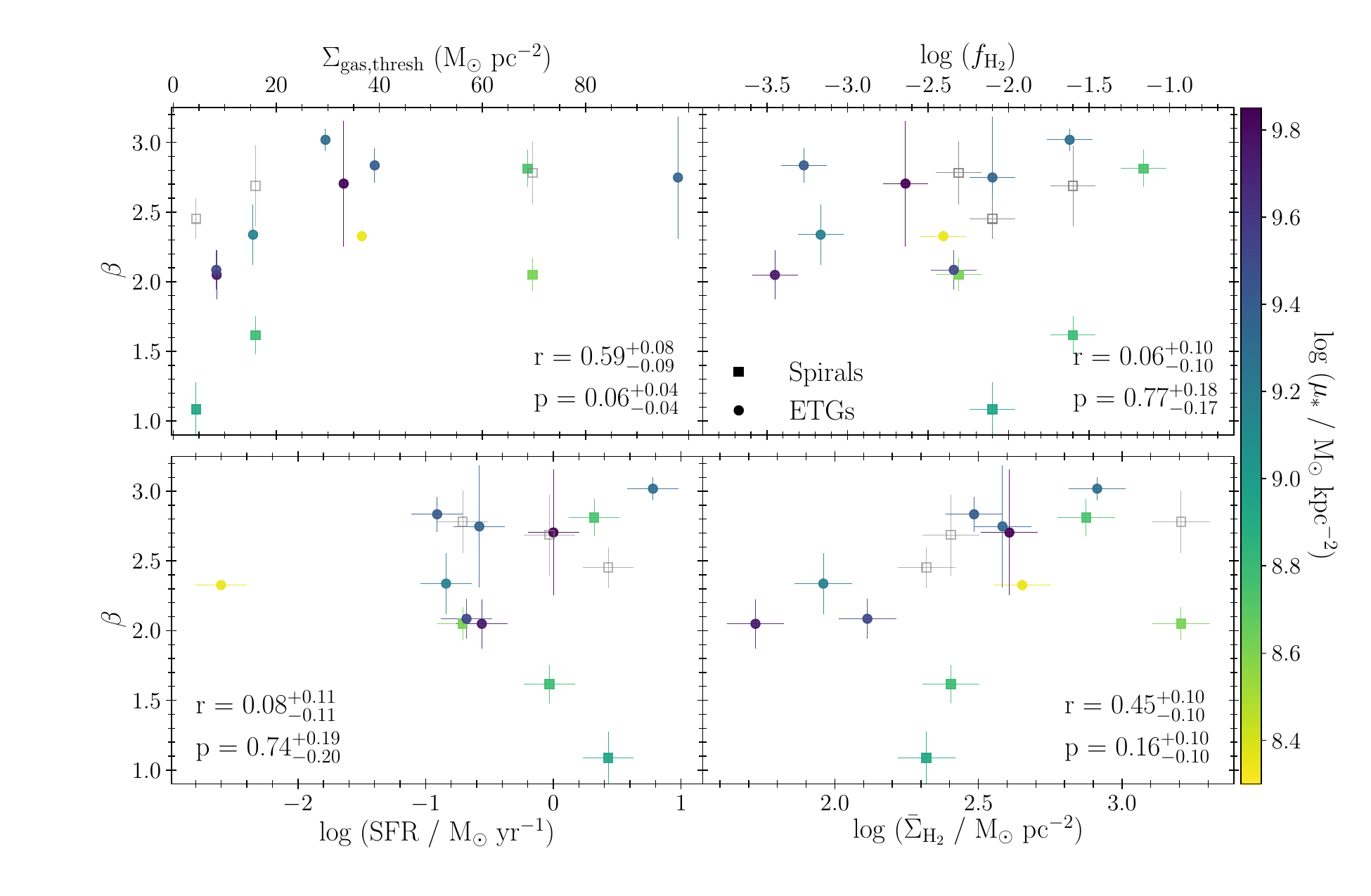}
    \caption{Slopes ($\beta$) of the power-laws best-fitting the power spectra of the WISDOM galaxies as a function of the galaxies' mass surface density sensitivities ($\Sigma_{\rm gas, thresh}$, top left), molecular gas fraction ($f_{\rm H_2}$, top right), SFR (bottom left) and average molecular gas mass surface density within the central kpc ($\bar{\Sigma}_{\rm H_2}$, bottom right), including the large scale slopes of the broken power-laws for NGC 4501, NGC 4826 and NGC 5806. The empty grey symbols show the result of the best-fitting single power-law to the aforementioned galaxies. The data points are colour-coded by their central stellar mass surface densities ($\mus$). Vertical error bars indicate the 1$\sigma$ uncertainties on the best-fitting power-law slopes and horizontal error bars show the uncertainty on the respective quantity. The Spearman rank correlation coefficient and p-value for each correlation are shown in each panel. There only statistically significant correlation is that between power spectrum slope and the average central molecular gas mass surface density of the observations.}
    \label{fig:pspec_obs_4panel_muscol_wlss}
\end{figure*}

Figures~\ref{fig:mus_slopes_wls} and~\ref{fig:pspec_obs_4panel_muscol_wlss} show the power spectrum slopes of either the best-fitting single power-laws or large scale slopes of the best-fitting broken power-laws of the WISDOM galaxies as a function of $\mus$ (Figure~\ref{fig:mus_slopes_wls}, cf. Figure~\ref{fig:pspec_slopes_mus_obs_sims}), surface density sensitivity, molecular gas-to-stellar mass ratio, star formation rate and average central molecular gas surface density (Figure~\ref{fig:pspec_obs_4panel_muscol_wlss}, cf. Figure~\ref{fig:pspec_obs_4panel_muscol}). The discrepancies between the trend between $\beta$ and $\mus$ of the simulated and observed galaxies are larger when considering the large scale slopes of the best-fitting broken power-laws. The large scale slopes are (much) shallower than the slopes of the best-fitting single power-law for the simulated galaxies in the same central stellar mass surface density regime. In contrast to the single power-law lopes, the combination of best-fitting single and large scale slopes shows a statistically significant correlation with the surface density sensitivity of the observations. 
There are still no trends between the best-fitting slopes and star formation rate or molecular gas-to-stellar mass fraction. However, while there still exists a putative trend of increasing power-law slope with central stellar surface density, it is no longer statistically significant (Spearman rank correlation coefficient $r = 0.45^{+0.10}_{-0.10}$ with a p-value $p = 0.16 ^{+0.10}_{-0.10}$) when considering the large scale slopes of the best-fitting broken power-laws for NGC 4501, NGC 4826 and NGC 5806. 

%%%%%%%%%%%%%%%%%%%%%%%%%%%%%%%%%%%%%%%%%%%%%%%%%%

% Don't change these lines
\bsp	% typesetting comment
\label{lastpage}
\end{document}